\RequirePackage[l2tabu, orthodox]{nag}
 \documentclass[11pt, reqno, letterpaper]{amsart}

\usepackage{graphicx, amsmath, amssymb, amsthm, mathtools, geometry, enumerate, color, bbm, bm, mathrsfs,multirow}
\usepackage[parfill]{parskip}
\usepackage{marginnote}
\usepackage[font=small, format=plain, labelfont=bf, up, textfont=it, up]{caption}
\usepackage{subcaption}
\usepackage[normalem]{ulem}
\usepackage{float}   
\usepackage{microtype}
\usepackage{tikz}
\usetikzlibrary{decorations.markings}
\usepackage[backend=bibtex,style=numeric]{biblatex}

\usepackage{cleveref}

\bibliography{FNLS}

\numberwithin{equation}{section}

\theoremstyle{plain} 
\newtheorem{theorem}{Theorem}[section]
\newtheorem{lemma}[theorem]{Lemma}
\newtheorem{prop}[theorem]{Proposition}
 
\theoremstyle{definition}
 
\theoremstyle{remark}
\newtheorem{remark}{Remark} [section]

\newtheoremstyle{RHP}  		
  {3pt}					
  {3pt}					
  {\itshape}				
  {}						
  {\bfseries}				
  {}						
  {.5em}					
 {\thmnote{#3}Riemann-Hilbert Problem #2 }

\theoremstyle{RHP}
\newtheorem{rhp}{}[section]

\newtheoremstyle{DBAR}  		
  {3pt}					
  {3pt}					
  {\itshape}				
  {}						
  {\bfseries}				
  {}						
  {.5em}					
  {\thmnote{#3}$\overline \partial$ Problem #2 } 

\theoremstyle{DBAR}
\newtheorem{DBAR}{}[section]

\newcommand{\R}{{\mathbb R}}

\newcommand{\C}{\mathbb{C}}

\newcommand{\La}{ {\mathcal{L} } }
\newcommand{\B}{ {\mathcal{B } } }

\newcommand{\dbar}{\overline{\partial}}
\newcommand{\pd}[3][ ]{\frac{\partial^{#1} #2}{\partial #3^{#1} } }

\renewcommand{\Re}{\mathop{ \mathrm{Re}}\nolimits}
\renewcommand{\Im}{\mathop{ \mathrm{Im} }\nolimits}
\newcommand{\im}{\mathrm{i} }

\newcommand{\bigo}[1]{\mathcal{O} \left( #1 \right)}
\newcommand{\littleo}[1]{ {o} \left( #1 \right) }

\newcommand{\lp}{\left(}
\newcommand{\rp}{\right)}
\newcommand{\lb}{\left[}
\newcommand{\rb}{\right]}

\newcommand{\vect}[1]{\bm{#1}}

\newcommand{\triu}[2][1]{\begin{pmatrix} #1 & #2 \\ 0 & #1 \end{pmatrix}}
\newcommand{\tril}[2][1]{\begin{pmatrix} #1 & 0 \\ #2 & #1 \end{pmatrix}}

\newcommand{\offdiag}[2]{\begin{pmatrix} 0 & #1 \\ #2 & 0 \end{pmatrix}}

\newcommand{\striu}[2][1]{\begin{psmallmatrix} #1 & #2 \\ 0 & #1 \end{psmallmatrix}}
\newcommand{\stril}[2][1]{\begin{psmallmatrix} #1 & 0 \\ #2 & #1 \end{psmallmatrix}}
\newcommand{\sdiag}[2]{\begin{psmallmatrix*}[r] #1 & 0 \\ 0 & #2 \end{psmallmatrix*}}
\newcommand{\soffdiag}[2]{\begin{psmallmatrix*}[r] 0 & #1 \\ #2 & 0 \end{psmallmatrix*}}

\newcommand{\U}{{\mathcal U}}
\newcommand{\one}{\bm{1}}

\newcommand{\ie}{i.e.}

\DeclareMathOperator*{\res}{Res}

\DeclareMathOperator{\sech}{sech}
\DeclareMathOperator{\dist}{dist}

\newcommand{\sig}{\sigma_3}
\newcommand{\poles}{\mathcal{Z}}

\newcommand{\data}{\sigma_d}
\newcommand{\eps}{\epsilon}
\newcommand{\mk}[1]{ M^{(#1)} }
\newcommand{\vk}[1]{ V^{(#1)} }
\newcommand{\Sk}[1]{ \Sigma^{(#1)} }
\newcommand{\Wk}[1]{ W^{(#1)} }
\newcommand{\indicator}{\chi_{_\poles}}

\newcommand{\Mrhp}{\mk{2}_{\textsc{rhp}}}

\newcommand{\mout}{M^{(\textrm{out})}}
\newcommand{\mxi}{M^{(\xi)}}
\newcommand{\mPC}{M^{(\textsc{pc})}}
\newcommand{\Ical}{\mathcal{I}}
\newcommand{\sol}{\mathrm{sol}}

\begin{document}

\title[Long time asymptotics of focusing NLS]{ Long time asymptotic behavior of the focusing Nonlinear Schrodinger equation}

\author{Michael Borghese}
\author{Robert Jenkins}
\author{Kenneth D. T.-R. McLaughlin}
\address{Department of Mathematics, University of Arizona, Tucson}
\email{mborghese@math.arizona.edu}
\email[Corresponding Author]{rjenkins@math.arizona.edu}
\email{mcl@math.arizona.edu}

\date{\today}

\begin{abstract}
We study the Cauchy problem for the focusing nonlinear Schr\"{o}dinger (NLS) equation. Using the $\dbar$ generalization of the nonlinear steepest descent method we compute the long time asymptotic expansion of the solution $\psi(x,t)$ in any fixed space-time cone $x_1 + v_1 t \leq x \leq x_2 + v_2 t$ with $v_1 \leq v_2$ up to an (optimal) residual error of order $\bigo{t^{-3/4}}$. In each $(x,t)$ cone the leading order term in this expansion is a multi-soliton whose parameters are modulated by soliton-soliton and soliton-radiation interactions as one moves through the cone. Our results only require that the initial data possess one $L^2(\R)$ moment and (weak) derivative and that it not generate any spectral singularities (embedded eigenvalues). 
\end{abstract}

\maketitle

\section{Introduction}\label{sec:intro}
In this paper we study the long time asymptotic behavior of the focusing nonlinear Schr\"{o}dinger (fNLS) equation on $\R \times \R_+$: 
\begin{equation}\label{nls}
	i \psi_t + \frac{1}{2} \psi_{xx} + | \psi |^2 \psi = 0, \qquad \psi(x, 0) = \psi_0(x).
\end{equation}

The long time behavior of the \textit{defocusing} NLS equation---equation \eqref{nls} with the sign of cubic nonlinearity reversed---has been thoroughly studied \cite{MZ,DIZ,DZ2,DZ5,DZ4,DM}. In the defocusing case, one finds that as $t \to \infty$, 
\begin{gather}\label{dnls longtime} 
	\psi(x,t) = t^{-1/2} \alpha(z_0) e^{i x^2/(2t) - i \nu(z_0) \log(4t)} + \mathcal{E}(x,t)
\shortintertext{where}
\nonumber
	\nu(z) = - \frac{1}{2\pi} \log (1 - |r(z)|^2), \quad 
	|\alpha(z)|^2 = \nu(z)^2,
\shortintertext{and}
\nonumber
	\arg \alpha(z) = \frac{1}{\pi} 
	\int_{-\infty}^z \log(z-s) d(\log(1-|r(s)|^2)) + \frac{\pi}{4} 
	+ \arg \Gamma(i \nu(z)) - \arg r(z).
\end{gather}
Here $z_0 = -x/(2t)$, $\Gamma$ is the gamma function, and $r$ is the so called reflection coefficient for the potential $\psi_0(x)$ described below.
Estimates for the size of the error term $\mathcal{E}(x,t)$ depend on smoothness and decay assumptions on $\psi_0$. The leading term without estimates was first obtained in \cite{MZ}. Using the nonlinear steepest descent method \cite{DZ3}, it was shown in \cite{DZ5,DZ2} that if $\psi_0$ had a high degree of smoothness and decay that $\mathcal{E}(x,t) = \bigo{ t^{-1} \log t}$. This was later improved \cite{DZ4} to $\mathcal{E}(x,t) = \bigo{ t^{-(1/2 + \kappa)} }$ for any $0 < \kappa < 1/4$ under the much weaker assumption that $\psi_0$ belonged to the weighted Sobolev space 
\[
	H^{1,1} = \left\{ f \in L^2(\R) \,:\, x f,\ f' \in L^2(R) \right\}.
\] 
Recently, McLaughlin and Miller \cite{MM1,MM2}, developed a method of asymptotic analysis of Riemann-Hilbert problems based on $\dbar$ problems, 
rather than the asymptotic analysis of singular integrals on contours. This was successfully adapted to study defocusing NLS both for finite mass initial data \cite{DM} and finite density initial data \cite{CJ}; the later of which supports soliton solutions. The advantages of this method are two fold: 1) it avoids delicate estimates involving $L^p$ estimates of Cauchy projection operators (central to the work in \cite{DZ4}), and 2) it improves error estimates without additional restrictions on the initial data. The result in \cite{DM}, which can be shown to be sharp, is that for $\psi_0 \in H^{1,1}$, the error $\mathcal{E}(x,t) = \bigo{t^{-3/4}}$.  

In this work we apply these $\dbar$-techniques to the inverse scattering transform (IST) for NLS to obtain the long-time asymptotic behavior of solutions to \eqref{nls}. The long-time behavior of solutions of focusing NLS are necessarily more detailed than in the defocusing case due to the presence of solitons which correspond to discrete spectra of the non self-adjoint ZS-AKNS (Dirac) scattering operator associated with focusing NLS (cf. \eqref{laxL} below). Given initial data $\psi_0 \in L^2(\R)$ the ZS-AKNS operator for \eqref{nls} allows for (complex conjugate pairs of) discrete spectra anywhere in $\C \backslash \R$. In the defocusing case the ZS-AKNS operator is self-adjoint and the discrete spectrum is empty for finite mass ($L^2(\R)$) initial data; discrete spectra are possible for the finite density type data studied in \cite{CJ}, but they are restricted to lie in a fixed interval of the real axis set by the initial data. The description of the minimal scattering data for the forward/inverse scattering transform is necessarily more complicated in the focusing case. 

Let us briefly consider the minimal scattering data for \eqref{nls}, more details are given in Section~\ref{sec:scattering} and the references therein. Associated with any $z_k \in \C^+$ of simple discrete spectrum is a nonzero complex number $c_k$ called a norming constant. The real axis is the continuous spectrum of the ZS-AKNS operator along which we define a \textit{reflection coefficient} $r:\R \to \C$. In the focusing case, the reflection coefficient $r$ may take any value in $\C$; it is also possible that $r$ may posses singularities along the real line---such points are called \textit{spectral singularities}. When spectral singularities exist it is possible for their to be a (countably) infinite discrete spectrum which must accumulate at a spectral singularity; if no spectral singularities exist, the discrete spectrum is finite. For initial data $\psi_0$ which produces only simple discrete spectra and has no spectral singularities, the minimal scattering data for focusing NLS is the collection $\mathcal{D} = \{ r(z), \{(z_k, c_k) \}_{k=1}^N \}$. This is the classical scattering map $\mathcal{S}: \psi_0 \mapsto \mathcal{D}$ for NLS. As described in \cite{BC,BDT} such initial data is generic. In the general, non-generic, case where spectral singularities or higher order spectra may exist the classical scattering map is replaced by $\mathcal{S}: \psi_0 \mapsto v$ where $v$ is a certain matrix defined along a contour $\Gamma$ consisting of the real axis and a closed circle around infinity as described in \cite{Zhou2}. 

In either case the amazing fact of integrability is that the scattering map $\mathcal{S}$ linearizes the time evolution; for a potential $\psi_0$ evolving according to \eqref{nls} the scattering data evolution is trivial: 
$\mathcal{D}(t) = \{ r(z) e^{2iz^2 t}, \{ (z_k, c_k e^{2i z_k^2 t} ) \}_{k=1}^N \}$ (or $v(t) = e^{-iz^2 t \sigma_3} v e^{iz^2 t \sigma_3}$ in the general case). 
It is often remarked in the literature that the scattering map $\mathcal{S}$ is a kind of nonlinear Fourier transform, and indeed it preserves regularity and smoothness in the same way; as shown in \cite{Zhou2} the scattering map is a bijective (in fact bi-Lipschitz) map from $H^{j,k}(\R)$ to $H^{k,j}(\Gamma)$ for any $j>0$ and $k \geq 1$ (in the classical setting without spectral singularities this reduces to the reflection coefficient $r \in H^{k,j}(\R)$). However, it is a trivial calculation that in order for the time evolving scattering data to persist in the weighted Sobolev space $H^{k,j}$ one must have $j \geq k$. It follows that the largest space $H^{j,k}$ from which the IST for \eqref{nls} is well defined in $H^{1,1}$, and this is precisely the space in which we will work.

Spectral data $\{ r\equiv 0, \{ (z_k, c_k)\}_{k=1}^N \}$ for which the reflection coefficient vanishes identically correspond to soliton solutions of \eqref{nls}. If the spectrum consist of a single point, $\sigma_d = \{(\xi + i \eta, c)\}$ the corresponding solution of \eqref{nls} is the one-soliton 
\begin{gather}
\begin{gathered} \label{one soliton}
	\psi_{\sol}( x, t) = \psi_{\sol} (x,t ; \{(\xi + i \eta, c) \} ) = 2\eta \sech( 2\eta (x + 2\xi t - x_0) )
		e^{-2i(\xi x + (\xi^2 - \eta^2)t)} e^{-i \phi_0}, \\
		\| \psi_{\sol}( \cdot, t) \|_{L^2(\R)}^2 = 4 \eta 
\end{gathered}
		\shortintertext{where the phase shift $x_0$ and constant $\phi_0$ are}
		x_0 = \frac{1}{2\eta} \log \left| \frac{ c}{2\eta} \right|,
		\qquad
		\phi_0 = \frac{\pi}{2} + \arg(c) \nonumber.
\end{gather}
This solution is a localized pulse with speed $v=-2\xi$ and maximum amplitude $2\eta$. 
When $N > 1$ the solution of \eqref{nls} with scattering data $\{ r\equiv 0, \ \sigma_d = \{ (z_k, c_k)\}_{k=1}^N \}$, which we label $\psi_\sol(x,t; \sigma_d)$, is called an $N$-soliton solution (corresponding to the discrete scattering data $\sigma_d$). The long time behavior of the $N$-soliton is a straightforward exercise in linear algebra and goes back to \cite{ZS}.
Generically, the solution  breaks apart into $N$ independent one-solitons; each traveling at distinct speed $v_k = -2\Re z_k$. When the spectra do not have distinct real parts the long-time behavior is more complicated;  
we give a streamlined review of this in Appendix~\ref{app:solitons}. 
Likewise, in the absence of solitons the defocusing methods mentioned above go through with only superficial changes of certain signs. The interesting question is how, in the generic case, the soliton and reflection coefficient terms interact to affect the long time limit. Formula \eqref{data modulation} used in Theorem~\ref{thm:main theorem} characterizes this interaction in the general setting and \eqref{limiting phases} shows explicitly how these interactions affect the asymptotic phase shifts of individual solitons.

\subsection{Main Results and Remarks}
Our main result describes the asymptotic behavior of the solution \eqref{nls} as $t \to \infty$, for generic initial data $\psi_0 \in H^{1,1}(\R)$. 
In order to state our results we define the following quantities derived from given scattering data $\{ r, \{(z_k, c_k) \}_{k=1}^N \}$. 
Let $\poles$ denote the projection of the discrete spectral data 
$\sigma_d = \{(z_k, c_k) \}_{k=1}^N$ onto its first coordinate $\poles = \{ z_k \}_{k=1}^N \subset \C^+$;
define 
\begin{gather}
	\nonumber
	\kappa(s) = - \frac{1}{2\pi} \log( 1 + |r(s)|^2),
	\intertext{and for any real number $\xi$ let}
	\nonumber
	\Delta_\xi^{-} = \{ k \in \{0,1, \dots, N \} \,:\, \Re z_k < \xi   \}.	
	\intertext{Given any real interval $\Ical = [a,b]$ let }
	\begin{gathered}\label{data modulation}
	\poles(\Ical) = \{ z_k \in \poles \,:\, \Re z_k \in \Ical \} 
	\quad \text{and} \quad N(\Ical) = | \poles(\Ical)| \\
	\Delta_\xi^{-}(\Ical) = 
	\{ k \in \{0,1, \dots, N \} \,:\, a \leq \Re z_k < \xi   \} 	\\
	\widehat{c}_k(\Ical) = c_k \prod_{\mathclap{
		\substack{
		z_j \in \poles \\
		\Re z_j < a } } }
		\Big( \frac{ z_k - z_j}{z_k - z_j^*}  \Big)^2
		\exp \lp \frac{i}{\pi} \int_{-\infty}^\xi 
		\log( 1 + |r(s)|^2) \frac{ds}{s-z_k} \rp
	\end{gathered}
\end{gather}
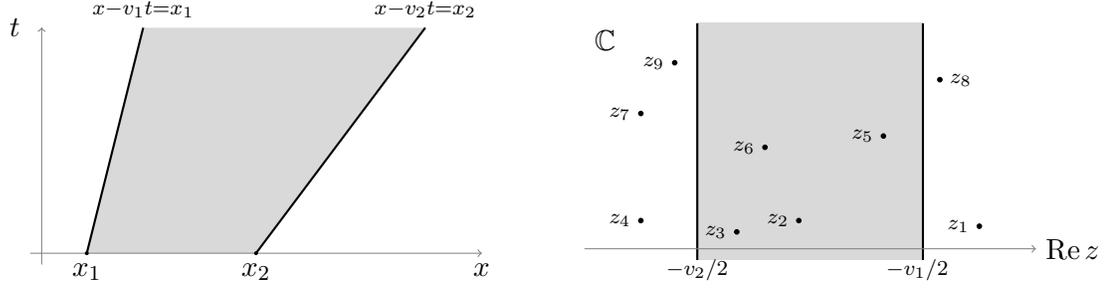
\begin{figure}[t]
\centering
\begin{tabular}{c c}
\begin{tikzpicture}[scale=0.75]
\path [fill=gray!30](-3,0) -- (-2,4) -- (3,4) -- (0,0)--(-3,0);
\draw [thick](-3,0) -- (-2,4);
\draw [thick](0,0) -- (3,4);
\draw [help lines][->] (-4,0) -- (4,0);
\draw [help lines][->] (-3.8,-0.2) -- (-3.8,4);
\draw [fill] (-3,0) circle [radius=0.025];
\draw [fill] (0,0) circle [radius=0.025];
\node [below] at (-3,0) {$x_1$};
\node [below] at (0,0) {$x_2$};
\node [below] at (4,0) {$x$};
\node [left] at (-4,4) {$t$};
\node [above] at (-2,4) {$\scriptstyle{x-v_1 t = x_1}$};
\node [above] at (3,4) {$\scriptstyle{x-v_2 t = x_2}$};
\end{tikzpicture}& 
\qquad
\begin{tikzpicture}[scale=0.75]
\path [fill=gray!30](-2,-0.2) -- (-2,4) -- (2,4) -- (2,-0.2)--(-2,-0.2);
\draw[thick](-2,-0.2) -- (-2,4);
\draw[thick](2,-0.2) -- (2,4);
\draw[help lines][->] (-4,0.0) -- (4,0.0);
\draw[fill] (3,.4) circle [radius=0.04];
\draw[fill] (2.3,3) circle [radius=0.04];
\draw[fill] (1.3,2) circle [radius=0.04];
\draw[fill] (-.2,.5) circle [radius=0.04];
\draw[fill] (-.8,1.8) circle [radius=0.04];
\draw[fill] (-1.3,.3) circle [radius=0.04];
\draw[fill] (-2.4,3.3) circle [radius=0.04];
\draw[fill] (-3,.5) circle [radius=0.04];
\draw[fill] (-3,2.4) circle [radius=0.04];
\node [left] at (3,.4) {$\scriptstyle{z_1}$};
\node [right] at (2.3,3) {$\scriptstyle{z_8}$};
\node [left] at (1.3,2) {$\scriptstyle{z_5}$};
\node [left] at (-.2,.5) {$\scriptstyle{z_2}$};
\node [left] at (-.8,1.8) {$\scriptstyle{z_6}$};
\node [left] at (-1.3,.3) {$\scriptstyle{z_3}$};
\node [left] at (-2.4,3.3) {$\scriptstyle{z_9}$};
\node [left] at (-3,.5) {$\scriptstyle{z_4}$};
\node [left] at (-3,2.4) {$\scriptstyle{z_7}$};
\node [below] at (-2,0) {$\scriptstyle{-v_2/2 }$};
\node [below] at (1.9,0) {$\scriptstyle{-v_1/2  }$};
\node [right] at (4,.0) {$\Re z$};
\node [right] at (-4,3.7) {$\C$};
\end{tikzpicture}\\ 
\end{tabular}
\caption{ Given initial data $\psi_0$ with scattering data $\{ r, \{(z_k, c_k)\}_{k=1}^N \}$, the asymptotic behavior of $\psi(x,t)$, the solution of \eqref{nls},  
in the space-time cone $x_1 + v_1 t \leq x \leq x_2+ v_2 t$ as $t\to \infty$ , is described to leading order by the $N(\Ical)$-soliton $\psi_\sol(x,t; \widehat{\sigma}_d)$ corresponding to the discrete spectral values in $\poles(\Ical)$ and connection coefficients $\widehat{c}_k$ modified by the self-interaction between solitons and with the reflection coefficient as described in Theorem~\ref{thm:main theorem}. 
In the example here, the original data has nine spectral values, but inside the shaded $(x,t)$ cone the solution is described by a 4-soliton with spectrum $\poles(\Ical) = \{ z_2, z_3, z_5, z_6\}$. 
\label{fig:cone to scattering}
}
\end{figure}
%
\begin{theorem}\label{thm:main theorem}
Let $\psi(x,t)$ be the solution of \eqref{nls} corresponding to initial data $\psi(x,t=0) = \psi_0(x) \in H^{1,1}(\R)$ and suppose that $\psi_0$ does not generate any spectral singularities. 
Let $\{ r, \{ z_k, c_k \}_{k=1}^N \}$ denote the spectral data generated from $\psi_0$. 
Fix $x_1, x_2, v_1, v_2 \in \R$ with 
$v_1 \leq v_2$. Let $\Ical = [-v_2/2, -v_1/2]$, and let $\xi = -x/(2t)$.
Then as $t \to \infty$ inside the truncated cone
\[
	x_1 + v_1 t \leq x \leq x_2+ v_2 t, \qquad t \to \infty
\]
we have
\begin{equation}
\psi(x,t)=\psi_\sol(x,t;\widehat{\sigma}_d ) + t^{-1/2}f(x,t)+\bigo{t^{-3/4}}.\nonumber
\end{equation}
Here, $\psi_\sol(x,t;\widehat{\sigma}_d)$ is the $N(\Ical)$ soliton corresponding to the modified discrete scattering data (see Figure~\ref{fig:cone to scattering}) given by 
$\widehat{\sigma}_d = \{ ( z_k, \widehat{c}_k(\Ical) ) \,:\, z_k \in \poles(\Ical) \}$,
with $\poles(\Ical)$ and $\widehat{c}_k(\Ical)$ as defined by \eqref{data modulation}, and 
\begin{equation}
f(x,t) = m_{11}(\xi)^{2}\alpha_{1}(\xi)e^{ix^{2}/(2t)-i\kappa(\xi)\log(4t)}+ m_{12}(\xi)^{2} \alpha_{2}(\xi)e^{-ix^{2}/(2t)+i\kappa(\xi)\log(4t)},\nonumber
\end{equation}
with
\begin{equation}
	|\alpha_{1}(\xi)|^{2} = |\alpha_{2}(\xi)|^{2} = |\kappa(\xi)|,
	\quad 
	\arg \alpha_{2}(s)=-\arg \alpha_{1}(s),\nonumber
\end{equation}
and
\begin{equation} 
\arg \alpha_{1}(s)=2\int_{-\infty}^\xi \frac{ \kappa(s) - \chi(s) \kappa(\xi)}{s-z} ds - 4 \sum\limits_{k \in \Delta_\xi^{-} } \arg(\xi - z_k) + \frac{\pi}{4}+\arg \Gamma(i\kappa(s))-\arg r(s).\nonumber
\end{equation}
The coefficients 
$m_{11}(\xi)$ and $m_{12}(\xi)$ 
are the entries in the first row of the solution of RHP~\ref{rhp:solitons2} with discrete spectral data $\widehat \sigma_d$ and $\Delta = \Delta_\xi^-(\Ical)$ evaluated at $z=\xi$.  
\end{theorem}


Our result is essentially optimal.  
For initial data in the weakest possible space in which the IST can be formulated, we derive an asymptotic description up to a residual $\bigo{t^{-3/4}}$ error; this is the same order that arises in the Fourier analysis of the free Schr\"{o}dinger equation $i\psi_t + \frac{1}{2} \psi_{xx} = 0$. 
We avoid the consideration of spectral singularities only to limit the length of the paper. Even subject to spectral singularities, our results should still hold in any $(x,t)$ cone $x_1 + v_1 t < x < x_2 + v_2 t$, such that the spectral interval $\mathcal{I}$ does not contain any spectral singularities. 

\begin{remark}
Spectral singularities may exist for data in any weighted Sobolev space $H^{j,k}$; there are even examples \cite[Example 3.3.16]{Zhou} of Schwarz class data for which spectral singularities occur. However, if the initial data decays exponentially, i.e., for some $c>0$, $\int_\R e^{c|x|} |\psi_0(x)|^2 dx < \infty$ then it is easily shown that spectral singularities cannot occur. 
\end{remark}

In Theorem~\ref{thm:main theorem} we give the asymptotic description in cones in order to accommodate many situations at once. In particular by considering small cones instead of fixed frames which of reference we are able to account for uncertainties in the computation (or measurement) of the spectral data and thus speed of the resulting solitons. We believe that such a description should also be useful to study non-integrable perturbations of focusing NLS where the discrete spectra would no longer be stationary. 

If one has additional knowledge of the spectral data, then the formulae above can be simplified greatly in fixed frames of reference $x - vt = \bigo{1}$.
In a frame of reference different than any soliton speed, i.e., if we have $|\xi - \Re z_k| \geq c >0$ for all $k=1,\dots,N$, then $\psi_\sol(x,t)$, $m_{11}(\xi)-1$, and $m_{12}(\xi)$ are each exponentially small in $t$ so the asymptotic description reduces to 
\[
	\psi(x,t) = t^{-1/2} \alpha_1(\xi) e^{i x^2/(2t) - i \kappa(\xi) \log(4t)}
	+\bigo{t^{-3/4}}.
\]
This is the analog of the defocusing result \eqref{dnls longtime}.
Next, we consider the frame of reference of a distinct 1-soliton, that is, suppose that $z_k = \xi_k + i \eta_k \in \poles$ is a discrete spectral value of the initial data $\psi_0$ whose real part is distinct from that of all other spectral values (except its complex conjugate) and let $c_k$ be its associated norming constant. Then 
as $t \to \infty$ with $x+2\Re(\xi_k) t = \bigo{1}$ the asymptotic solution reduces to
\begin{subequations}\label{limiting phases}
\begin{equation}
\begin{gathered}
	\psi(x,t) = \psi_\sol(x,t;(z_k, \widehat c_k))  + \bigo{t^{-1/2}}  \\
	\psi_\sol(x,t;(z_k,\widehat c_k))  = 
	2\eta_k \sech\lp 2\eta_k(x+2\xi_k t -x_0) \rp 
	e^{-2i( \xi_k x + (\xi_k^2-\eta_k^2)t ) } e^{-i \phi_0}
\end{gathered}
\end{equation}
where
\begin{equation}
\begin{gathered}
	x_0 = \frac{1}{2\eta_k} \log \left| \frac{c_k}{2\eta_k} \right|
		+\eta_k^{-1} \sum_{\mathclap{\ \ \Re z_j < \xi_k} }
			\log \left| \frac{ z_k - z_j}{z_k - z_j^*} \right|
		-\frac{1}{2\pi} \int\limits_{-\infty}^{\mathclap{-x/(2t)}} 
		\log(1+|r(s)|^2) \frac{ds}{(s-\xi_k)^2+\eta_k^2}	 \\
	\phi_0 = \frac{\pi}{2} + \arg c_k 
	 + 2 \sum_{\mathclap{\ \ \Re z_j < \xi_k} }
			\arg \lp \frac{ z_k - z_j}{z_k - z_j^*} \rp
		+\frac{1}{\pi} \int\limits_{-\infty}^{\mathclap{-x/(2t)}} 
		\log(1+|r(s)|^2) \frac{s-\xi_k}{(s-\xi_k)^2+\eta_k^2} ds
\end{gathered}
\end{equation}
\end{subequations}
describe the asymptotic phase shifts. The last two terms in each expression above describe the asymptotic effect of the soliton-soliton interaction and the interaction of the soliton with the radiative component of the solution respectively. 

If all of the solitons have distinct real parts, then the solution separates asymptotically in the sense that uniformly for $x\in \R$ as $t \to \infty$,
\[
	\psi(x,t) = \sum_{k=1}^N \psi_\sol(x,t; (z_k, \widehat c_k) ) + \bigo{t^{-1/2}},
\]
and the correction of order $t^{-1/2}$ can be explicitly computed using the results of Theorem~\ref{thm:main theorem}.

\begin{remark}
Though we say that initial data $\psi_0$ whose spectra have distinct real parts are generic (in the sense that small perturbations of any non-generic initial datum will be generic) there are important classes of non-generic data. The so called Klaus-Shaw `single lobe' potentials, $\psi_0(x) = A(x)e^{i k x + i \phi_0}$ with $k,\phi_0 \in \R$ and $A(x)$ a bounded piecewise smooth  function which is nondecreasing to the left of some $x_0$ and nonincreasing to the right of $x_0$, are such that all of the discrete spectra have the same real part. Such potentials have been extensively studied in the semi-classical limit where the number of spectra is asymptotically large.   
\end{remark}

\subsection*{Organization of the rest of the paper}
In Section~\ref{sec:scattering} we describe the forward scattering transform step of the IST in greater detail collecting the necessary results for our later work and provide references for their proofs. The section ends with the characterization of the inverse scattering transform in terms of a Riemann-Hilbert problem RHP~\ref{rhp:M}. Section~\ref{sec:conj} begins the Riemann-Hilbert analysis by describing the initial conjugation of RHP~\ref{rhp:M} to better condition the problem for asymptotic analysis in a given frame of reference. Section~\ref{sec:extensions} introduces the $\dbar$ analysis to define extensions of the jump matrix for the non-linear steepest descent method. In Section~\ref{sec:models} we construct a global model solution which captures the leading order asymptotic behavior of the solution. Removing this component of the solution results in a small-norm $\dbar$ problem which is analyzed in Section~\ref{sec:dbar} culminating in a proof of our main result Theorem~\ref{thm:main theorem}.

\section{Results of scattering theory for focusing NLS}\label{sec:scattering}

The focusing NLS equation can be integrated \cite{AKNS,ZS} using the ZS-AKNS operator associated with Lax pair for NLS:
\begin{subequations}\label{lax pair}
	\begin{align}
	\label{laxL}
		&(\partial_x  - \La) \Phi = 0, 
		\qquad 
		\La = -i z \sigma_3 + \Psi, \\ 
	\label{laxB}
		&(i\partial_t - \B) \Phi = 0, 
		\qquad 
		\B = iz \mathcal{L} + \frac{1}{2} \sigma_3( \Psi^2 - \Psi_x),
\end{align}
where
\[
	\Psi = \Psi(x,t) = \offdiag{ \psi(x,t) } {-\psi(x,t)^*},
\]
\end{subequations}
and $\sigma_3$ is the third Pauli matrix
$
	\sigma_1 = \soffdiag{1}{1}, 
	\
	\sigma_2 = \soffdiag{-i}{i},
	\
	\sigma_3 = \sdiag{1}{-1}.
$		
The existence of a simultaneous solution of this overdetermined system  of equations requires that the potential $\Psi(x,t)$ satisfy the zero-curvature equation,
\begin{equation}
	i \mathcal{L}_t - \mathcal{B}_x + [ \, \mathcal{L}, \, \mathcal{B} \, ]
	= i \Psi_t + \frac{1}{2}\sig \Psi_{xx} - \sig \Psi^2 = 0,
\end{equation}
which is just a restatement of \eqref{nls}.


In the forward scattering step given initial data $\psi_0(x)$ one constructs solutions $\Phi(x,z)$ of \eqref{laxL} with $z \in \R$; in particular one constructs the two \textit{Jost solutions} 
$\Phi^{(\pm)}(x;z) = m^{(\pm)}(x,z)e^{-izx \sig}$, which satisfy 
\begin{equation}\label{jost diff eq}
	i \partial_x m = - i z [ \sig, m] + \Psi m, \qquad 
	\lim_{x \to \pm \infty} m^{\pm}(x,z) = I.
\end{equation}
These solutions can be expressed as Volterra type integrals
\[
	m^{(\pm)}(z) = I + \int_{\pm \infty}^x 
	e^{iz(x-y) \sig} \Psi(y) m^{(\pm)}(y) e^{-iz(x-y) \sig} dy
\]
By iteration one shows that these equations have bounded continuous solutions in both $x$ and $z$ whenever $\psi_0 \in L^1(\R)$. 

As the differential equation \eqref{laxL} is traceless, the determinant of any solution $\Phi$ is independent of $x$ and it follows that $\det \Phi^{(\pm)} = m^{(\pm)} \equiv 1$; for complex $z$ solutions must also possess the symmetry 
$m(x, z^*) = \sigma_2 m(x,z)^* \sigma_2$. 
It follows that for $z \in \R$ both $m^{(+)}$ and $m^{(-)}$ define a fundamental solution set for \eqref{jost diff eq} and so there exists a continuous matrix function $S(z)$, \textit{the scattering matrix}, satisfying
\begin{equation}\label{scattering matrix}
	\begin{gathered}
	\Phi^{(-)}(x;z) = \Phi^{(+)}(x;z) S(z), \qquad z \in \R, \\
	S(z) = \begin{pmatrix} a(z) & -b(z)^* \\ b(z) & a(z)^* \end{pmatrix},
	\qquad
	\det S(z) = |a(z)|^2 + |b(z)|^2 = 1
	\end{gathered}
\end{equation}
the coefficients $a(z)$ and $b(z)$ can be expressed as 
\begin{equation}
	\begin{aligned}
	a(z) &= \det \lb m_1^{(-)}, \, m_2^{(+)} \rb
		= 1 + \int_R \psi(y)^* m_{12}^{(+)} (y) dy
		= 1 + \int_R \psi(y)   m_{21}^{(-)} (y) dy,  \\
	b(z) &= \det \lb m_1^{(+)}, \, m_1^{(-)} \rb
		= - \int_R \psi(y)^* e^{-2izy} m_{11}^{(+)}(y) dy
		= - \int_R \psi(y)^* e^{-2izy} m_{11}^{(-)}(y) dy
	\end{aligned}
\end{equation}
where
\[
	m^{(\pm)} 
	= \begin{pmatrix} m^{(\pm)}_{1} \,, \, m^{(\pm)}_{2} \end{pmatrix} 
	= \begin{pmatrix} 
		m^{(\pm)}_{11} & m^{(\pm)}_{12} \smallskip \\
		m^{(\pm)}_{21} & m^{(\pm)}_{22}
	\end{pmatrix}.
\]

The following results are standard, proofs and details can be found in the literature, see for example \cite{BC,BDT,DZ4}.

Let $m^{(\pm)}_j$ denote the $j^{th}$ column of $m^{(\pm)}$ and $e_j$ denote the $j^{th}$ column of the identity matrix: 
\begin{itemize}
	\item $m^{(-)}_1(x,z)$, $m^{(+)}_2(x,z)$ and $a(z)$ extend analytically to $z \in \C^+$ with continuous boundary values on $\R$. 
	As $z \to \infty$ in $\C^+$, $m^{(-)}_1(x,z) \to e_1$, $m^{(+)}_2(x,z) \to e_2$ and $a(z) \to 1$.  
	Analogous statements hold for the other pair of columns for $z \in \C^-$. 
	Generally, $b(z)$ is defined only for $z \in \R$.
	
	\item At any $z_k \in \C^+$ for which $a(z_k) = 0$, the	solutions $\Phi_1^{(-)}(x,z_k)$ and $\Phi_2^{(+)}(x,z_k)$ are linearly dependent. 
	Specifically, a \textit{norming constant} $c_k$ exists such that:
	\[
		\Phi_1^{(-)}(x,z_k) = c_k \Phi_2^{(+)}(x,z_k).
	\] 
	As these solution decays exponentially as $x \to \mp \infty$ respectively, 
	this indicates that $z_k$ is an $L^2$ eigenvalue of \eqref{laxL} with eigenfunction $\Phi_1^{(-)}(x;z_k)$. 
	The symmetry $a(z^*) = a(z)^*$ implies these eigenvalues come in conjugate pairs.
	
	
	\item The \textit{reflection coefficient} $r:\R \to \C$ and \textit{transmission coefficient} $\tau:\C^+ \to \C$ are defined by
	\begin{equation}
		r(z) = \frac{b(z)}{a(z)} \qquad \qquad 
		\tau(z) = \frac{1}{a(z)}
	\end{equation}
	and it follows from \eqref{scattering matrix} that $1+ |r(z)|^2 = |\tau(z)|^2$ for each $z\in \R$.

	\item The properties of the scattering coefficients are similar to those of the Fourier transform. Given initial data $\Psi_0$ in the weighted Sobolev space
	\[
		H^{j,k}(\R)  = \left\{ f \in L^2(\R) \,: 
		\partial_x^j f ,\, |x|^{k} f \in L^2(\R) \right\}
	\]
	the scattering coefficients $a(z)-1 \in H^{k,1}$ and $b(z) \in H^{k,j}$. 
	It follows that, in the absence of spectral singularities (real zeros of $a(z)$), the map $\mathcal{R}:\psi_0 \mapsto r$ is a map from $H^{j,k}$ to $H^{k,j}$ (c.f. \cite{DZ4}).
\end{itemize}

%
%
%
%
%
%
%
%

The collection of data $\mathcal{D} = \{ r(z), \{z_k, c_k\}_{k=1}^N \}$ are called the scattering data for $\psi_0(x)$ and the map $ \mathcal{S}:  \psi_0 \mapsto \mathcal{D}$ is called the (forward) scattering map. The essential fact of integrability, is that if the potential $\psi_0(x)$ evolves according to \eqref{nls} then the evolution of the scattering data $\mathcal{D}$ is trivial
\begin{equation}
	\mathcal{D}(t) 
	= \left\{ r(z,t), \{ z_k(t), c_k(t) \}_{k=1}^N \right\} 
	= \left\{ r(z) e^{2it z^2}, \{ z_k , c_k e^{2it z_k^2} \}_{k=1}^N \right\}.
\end{equation}
The inverse scattering map $\mathcal{S}^{-1}: \mathcal{D}(t) \mapsto \psi(x,t)$ seeks to recover the solution of \eqref{nls} from its scattering data. 
This is done as follows:
from the (now time evolving) Jost function 
$\Phi^{(\pm)}(x,t;z) = m^{(\pm)}(x,t;z) e^{-izx \sig}$ 
one constructs the function 
\begin{equation}\label{M fscat}
	 M(z) = M(z;x,t) := \left\{ \begin{array}{c@{\quad:\quad}l}
	 	\begin{bmatrix} 
		\dfrac{ m^{(-)}_1(x,t;z)}{a(z)} \, , \, m_2^{(+)}(x,t;z) 
		\end{bmatrix} \smallskip
		& z \in \C^+ \\
		\sigma_2 M(z^*;x,t)^* \sigma_2 & z \in \C^-.
	\end{array} \right.
\end{equation}

Assuming that the data $\psi_0$ is generic in the sense that $a(z)$ has only simple zeros in $\C^+$ and no spectral singularities, the matrix $M$ defined above is the solution of the following Riemann-Hilbert problem. 
\begin{rhp}\label{rhp:M}
Find a meromorphic function $M: \C \backslash (\R \cup \poles \cup \poles^*) \to SL_2(\C)$ with the following properties
\begin{enumerate}[1.]
	\item $M(z) = I + \bigo{z^{-1}}$ as $z \to \infty$. 
	\item For each $z \in \R$, $M$ takes continuous boundary values
	$M_\pm(z) := \lim_{\eps \to 0^+} M(z \pm \im \eps)$ 
	which satisfy the jump relation $M_+(z) = M_-(z) V(z)$ where
	\begin{equation}\label{V}
		V(z) = \begin{pmatrix} 1 + |r(z)|^2 & r^*(z) e^{-2it\theta(z)}\\ r(z)e^{2i t \theta(z)} & 1 \end{pmatrix},
	\end{equation}
	where 
	\begin{equation}\label{phase}
		\theta = \theta(z;x,t) = z^2 - 2 \xi z = (z-\xi)^2 - \xi^2, \quad \xi = -x/(2t) .
	\end{equation}
	\item $M(z)$ has simple poles at each $z_k \in \poles$ and $z_k^* \in \poles^*$ at which 
	\begin{equation}\label{residues}
		\begin{aligned}
            		\res_{z_k} M &= \lim_{z \to z_k} M 
            		\begin{pmatrix} 0 & 0 \\ c_k e^{2it \theta} & 0 \end{pmatrix}, \\
            		\res_{z_k^*} M &= \lim_{z \to z^*_k} M 
            		\begin{pmatrix} 0 & -c^*_k e^{-2it \theta} \\ 0 & 0 \end{pmatrix}.
		\end{aligned}
	\end{equation}
\end{enumerate}
\end{rhp}
It's a simple consequence of Liouville's theorem that if a solution exists it is unique.
Expanding this solution as $z \to \infty$, $M = I + z^{-1} M^{(1)}(x,t) + \littleo{z^{-1}}$ and inserting this into \eqref{jost diff eq} one finds that
\begin{equation}\label{M expand}
	M = I + \frac{1}{2i z} 
		\begin{bsmallmatrix} 
		- \int_{x}^\infty |\psi(s,t)|^2 ds &
		\psi(x,t) \\
		\psi(x,t)^* &
		 \int_{x}^\infty |\psi(s,t)|^2 ds 
		\end{bsmallmatrix}
		+ \littleo{z^{-1}},
\end{equation}
and it follows that the solution of \eqref{nls} is given by 
\begin{equation}\label{recover}
	\psi(x,t) = \lim_{z \to \infty} 2i z M_{12}(z;x,t).
\end{equation}

For non-generic potentials various parts of the above characterization must be altered. 
There can exist points $z \in \R$ for which $a(z) = 0$, in which case $m_\pm(x,z)$ fail to exist; these are called \emph{spectral singularities}. 
The number of discrete spectra of \eqref{laxL} may be infinite, due to the first property of the solution $m$, the discrete spectra must accumulate at a spectral singularity along the real axis. 
In the absence of spectral singularities the discrete spectrum is finite. 
Smoothness and decay of the initial data does not preclude the existence of spectral singularities; in \cite{Zhou} an explicit example is given of a Schwarz-class potential which generates an infinite discrete spectrum accumulating at $z=0$. 
Finally, even in the case of a finite spectrum, poles may coalesce resulting in higher order singularities at certain points of the discrete spectrum; in this case the pole conditions \eqref{residues} must be altered.  
For simplicity we will consider here only the generic setting. 
Special cases of a single spectral singularity and of an infinite number of solitons have been partially described in \cite{Kam1,Kam2}.

%
%
%
%

\section{Conjugation}\label{sec:conj}
The function $M(z;x,t)$ defined by \eqref{M fscat} which solves solvs RHP~\ref{rhp:M}, is normalized such that it has identity asymptotics as $x \to +\infty$ with $t$ fixed. It is not unreasonable to assume that the RHP should be well conditioned as $t\to \infty$ along a characteristic $x = vt$ where $v \gg 1$. However, along an arbitrary characteristic there is no reason to expect that $M$ will remain near identity. In this section we describe a transformation $M \mapsto \mk{1}$ which renormalizes the RHP such that it is well behaved as $t \to \infty$ along an arbitrary characteristic. 
 
Let $\xi = -x/(2t)$. Define the partition of $\{ 0,1, \dots, N\} = \Delta_{\xi}^- \cup \Delta_\xi^+$ by 
\begin{equation}\label{index sets}
	\begin{aligned}
		\Delta_\xi^{-} &= \{ k \in \{0,1, \dots, N \} \,:\, \Re z_k < \xi   \}, \\
		\Delta_\xi^{+} &= \{ k \in \{0,1, \dots, N \} \,:\, \Re z_k \geq \xi   \}.
	\end{aligned}
\end{equation}
This partition splits the the residues $c_k$ in \eqref{residues} into two sets:
As $t \to \infty$ with $x \geq -2 \xi t$, 
it follows from \eqref{phase} that for each $k \in \Delta_\xi^-$, $\Im( \theta(z_k)) < 0$ and thus the residue of $M(z)$ at $z_k$ in \eqref{residues} grows without bound as $t \to \infty$, similarly, for $z_k$ with $k \in \Delta_\xi^+$, the resides are bounded or near zero. 
 
The first step in our analysis is to introduce a transformation which renormalizes the Riemann-Hilbert problem such that it is well conditioned for $t \to \infty$ with $\xi$ fixed. 
In order to arrive at a problem which is well normalized, we introduce the function
\begin{equation}\label{trans}
	\begin{gathered}
	T(z) = T(z, \xi) = 
	\prod_{k \in \Delta^{-}_\xi} \lp \frac{ z - z_k^*}{z-z_k}  \rp 
	\exp \lp i \int_{-\infty}^\xi  \frac{\kappa(s) }{s-z} ds\rp,
	\\
	\kappa(s) = -\frac{1}{2\pi} \log(1 + |r(s)|^2).
	\end{gathered}
\end{equation}
A standard result of the forward scattering theory \cite{FT} is the following trace formula for the transmission coefficient 
\begin{equation}\label{trace formula}
	\frac{1}{a(z)} = 
	\prod_{k =1}^N \lp \frac{ z - z_k^*}{z-z_k}  \rp 
	\exp \lp  \frac{1}{2\pi i} \int_{-\infty}^{\infty} \log(1 + |r(s)|^2) \frac{ds}{s-z} \rp  
\end{equation}
from which we see that our function $T(z,\xi)$ is a partial transmission coefficient which approaches the total transmission $1/a(z)$ as $\xi \to \infty$.  	
\begin{prop}\label{prop:T}
The function $T(z)$ defined by \eqref{trans} has the following properties:
\begin{enumerate}[a.]
	\item $T$ is meromorphic in $\C \backslash (-\infty, \xi]$. For each $k \in \Delta_\xi^-$, 
	$T(z)$ has a simple pole at $z_k$ and a simple zero at $z_k^*$.
	\item For $z \in \C \backslash (-\infty, \xi]$, $T(z^*)^* = 1/T(z)$.
	\item For $z \in (-\infty, \xi)$, the boundary values $T_\pm$ satisfy 
	\begin{equation}\label{Tjump}
		T_+(z) / T_-(z) = 1 + |r(z)|^2, \quad z \in (-\infty, \xi).
	\end{equation}
	
	\item As $|z| \to \infty$ with $|\arg(z)| \neq \pi$, 
	\[
		T(z) = 1 + \frac{i}{z} \Bigg[  2 \sum_{k \in \Delta^{-}_{\xi} } \Im z_k 
		- \frac{1}{2\pi} \int_{-\infty}^\xi \log( 1 + |r(s)|^2 ) ds \Bigg]  
		+ \bigo{ z^{-2} }.
	\]
	\item As $z \to \xi$ along any ray $\xi + e^{i \phi} \R_+$ with $| \arg \phi|< \pi$
	\begin{gather} \label{Tbound}
		\left| T(z,\xi) - T_0(\xi) (z-\xi)^{i \kappa(\xi)} \right| 
		\leq C \| r \|_{H^1(\R)} |z - \xi |^{1/2}
	\shortintertext{where $T_0(\xi)$ is the complex unit}
		\nonumber
		T_0(\xi) =\prod_{k \in \Delta^{-}_\xi} \lp \frac{ \xi - z_k^*}{\xi - z_k} \rp
		e^{i \beta(\xi,\xi) }
		= \exp \lb 
		i \lp \beta(\xi,\xi) - 2 \sum\limits_{k \in \Delta_\xi^{-} } \arg(\xi - z_k) 
		\rp \rb,
		\\ \nonumber
		\beta(z,\xi) = - \kappa(\xi) \log(z-\xi+1)
		+ \int_{-\infty}^\xi \frac{ \kappa(s) - \chi(s) \kappa(\xi)}{s-z} ds
		,
	\end{gather}
	and $\chi(s)$ is the characteristic function of the interval $(\xi-1, \xi)$ and the logarithm is principally branched along $(-\infty,\xi-1]$. 
\end{enumerate}
\end{prop}

\begin{proof}
Parts $a$.--$c$. are elementary consequences of the definition \eqref{trans} and the Sokhotski-Plemelj formula. 
For part $d$. one geometrically expands the product term and the factor $(s-z)^{-1}$ for large $z$, and use the fact that $\| \kappa \|_{L^1(\R)} \leq \| r \|_{L^2(\R)}$ to bound the remainder in the integral term for $z$ bounded away from the contour of integration. 
For part $e$. we write
\[
	\begin{aligned}
	T(z,\xi) 
	&= \prod_{k \in \Delta^{-}_\xi} \lp \frac{z- z_k^*}{z-z_k} \rp 
	\exp \lp 
 	i \int_{\xi-1}^\xi \frac{\kappa(\xi)} {s-z} ds  
	+ i \int_{-\infty}^\xi \frac{ \kappa(s) - \chi(s) \kappa(\xi)} {s-z} ds 
	\rp \\
	&= \prod_{k \in \Delta^{-}_\xi} \lp \frac{z- z_k^*}{z-z_k} \rp 
	(z- \xi)^{i \kappa( \xi)} \exp \lp i \beta(z,\xi) \rp.
	\end{aligned}
\]  
The result then follows from the facts that 
$\left| (z-\xi)^{i \kappa(\xi)} \right| \leq e^{-\pi \kappa(\xi)} 
= \sqrt{ 1 + |r(\xi)|^2}$ and using Lemma 23.3 of \cite{BDT} 
\[
	\left| \beta(z,\xi) - \beta(\xi, \xi) \right| 
	\leq C \| r \|_{H^1(\R)} |z - \xi|^{1/2}.
\]
\end{proof}
We define a new unknown function $\mk{1}$ using our partial transmission coefficient 
\begin{equation}\label{M1}
\mk{1}(z) = M(z) T(z)^{-\sigma_3}
\end{equation}

\begin{prop}\label{prop:M1}
The function $\mk{1}$ defined by \eqref{M1} satisfies the following Riemann-Hilbert problem
\begin{rhp}\label{rhp:M1}
Find a meromorphic function $\mk{1}: \C \backslash \R  \to SL_2(\C)$ with the following properties
\begin{enumerate}[1.]
	\item $\mk{1}(z) = I + \bigo{z^{-1}}$ as $z \to \infty$. 
	\item For each $z \in \R$, the boundary values
	$\mk{1}_\pm(z)$ satisfy the jump relation $\mk{1}_+(z) = \mk{1}_-(z) \vk{1}(z)$ where
	\begin{equation}\label{V1}
		\vk{1} (z) = \begin{cases}
			\triu{ r^*(z) T(z)^{2} e^{-2it \theta} }
			\tril{ r(z) T(z)^{-2} e^{2it \theta} }	
			& z \in (\xi, \infty) \\[1.5em]
			\tril{ \frac{ r(z) T_-(z)^{-2}}{1+ |r(z)|^2} e^{2it \theta} }
			\triu{ \frac{ r^*(z) T_+(z)^{2} }{1+ |r(z)|^2} e^{-2it \theta} }
			& z \in (-\infty, \xi)
		\end{cases}
	\end{equation}
	\item $\mk{1}(z)$ has simple poles at each $z_k \in \poles$ and $z_k^* \in \poles^*$ at which 
	\begin{equation}\label{residues1}
		\begin{aligned}
			\res_{z_k} \mk{1} &= 
			\begin{dcases}
				\lim_{z \to z_k} \mk{1}
            			\triu[0]{ c_k^{-1} (1/T)'(z_k)^{-2} e^{-2it \theta} } & k \in \Delta_\xi^{-} \\
				\lim_{z \to z_k} \mk{1}
				\tril[0]{ c_k T(z_k)^{-2} e^{2it \theta} } & k \in \Delta_\xi^{+} 
			\end{dcases} \\
            		\res_{z_k^*} \mk{1} &= 
			\begin{dcases}
        				\lim_{z \to z^*_k} \mk{1} 
                    		\tril[0]{ -(c_k^*)^{-1} T'(z_k^*)^{-2} e^{2it \theta} } & k \in \Delta_\xi^{-}  \\
        				\lim_{z \to z^*_k} \mk{1} 
                    		\triu[0]{ -c^*_k T(z_k^*)^{2} e^{-2it \theta} } & k \in \Delta_\xi^{+} 
			\end{dcases}
		\end{aligned}
	\end{equation}
\end{enumerate}
\end{rhp}
\end{prop}

\begin{proof}
That $\mk{1}$ is unimodular, analytic in $\C \backslash (\R \cup \poles \cup \poles^*)$, and approaches identity as $z \to \infty$ follows directly from it's definition, Proposition~\ref{prop:T} and the properties of $M$. The jump \eqref{V1} follows directly from using the factorizations of $V$, \eqref{V}, given by 
\begin{equation*}
	\vk{1}(z)= \begin{cases}
	T(z)^{\sigma_3} 
	\triu{r(z)^* e^{-2it \theta} } \tril{r(z) e^{2it \theta} }  
	T(z)^{-\sigma_3} 
	&  z > \xi \\
	T_-(z)^{\sigma_3} \tril{ \frac{r(z) e^{2it \theta} }{1+ |r(z)|^2}  } 
	\lp \dfrac{T_+(z) }{ T_-(z)} \rp^{\sigma_3} 
	\triu{ \frac{r(z)^* e^{-2it \theta} }{1+|r(z)|^2}  }
	T_+(z)^{-\sigma_3}
	& z < \xi
	\end{cases}
\end{equation*}
to the right and left of $z= \xi$ on the real line respectively and making use of the jump relation \eqref{Tjump} satisfied by $T(z)$ on $(-\infty, \xi)$. Concerning the residues, since $T(z)$ is analytic at each $z_k, z_k^*$ with $k \in \Delta^+_\xi$,  the residue conditions at these poles are an immediate consequence of \eqref{trans}. 
For $k \in \Delta^-_\xi$, $T(z)$ has a zero at $z_k^*$ and a pole at $z_k$, so that $\mk{1}_1 = M_1(z) T(z)^{-1}$ has a removable singularity at $z_k$, but acquires a pole at $z_k^*$. For $\mk{1}_2 = M_2(z) T(z)$ the situation is reversed; it has a pole at $z_k$ and a removable singularity at $z_k^*$. At $z_k$ we have
\[
	\begin{gathered}
	\begin{aligned}
		\mk{1}_1(z_k) &= \lim_{z \to z_k} M_1(z) T(z)^{-1} = \res_{z_k} M_1(z)  \cdot (1/T)'(z_k) \\
		&= c_k e^{2it\theta_k} M_2(z_k)  (1/T)'(z_k),  
	\end{aligned} \\
	\begin{aligned}
		\res_{z_k} \mk{1}_2(z) &= \res_{z=z_k} M_2(z) T(z) = M_2(z_k) \lb (1/T)'(z_k) \rb^{-1}  \\
		&= c_k^{-1} \lb (1/T)'(z_k) \rb^{-2} e^{-2it \theta} \mk{1}_1(z_k), 
	\end{aligned}
	\end{gathered}
\]
from which the first formula in \eqref{residues1} clearly follows. The computation of the residue at $z_k^*$ for $k \in \Delta^-_\xi$ is similar. 
\end{proof}

\section{Introducing \texorpdfstring{$\dbar$}{DBAR} extensions of jump factorization}\label{sec:extensions}

The next step in our analysis is to introduce factorizations of the jump matrix whose factors admit continuous--but not necessarily analytic--extensions off the real axis. Using these extensions we define a new unknown that deforms the oscillatory jump along the real axis onto new contours along which the jumps are decaying. The price we pay for this non-analytic transformation is that the new unknown has nonzero $\dbar$ derivatives inside the regions in which the extensions are introduced and satisfies a hybrid $\dbar$/Riemann-Hilbert problem. 

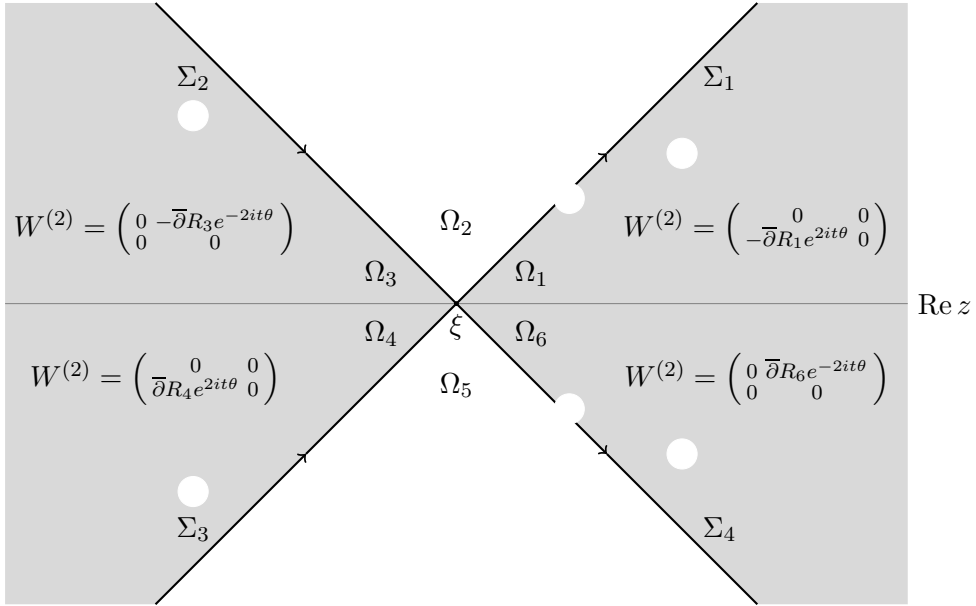
\begin{figure}[t]
\centering
\begin{tikzpicture}
\path [fill=gray!30] (0,0) -- (-4,4) -- (-6,4) -- (-6,-4) -- (-4,-4) -- (0,0);
\path [fill=gray!30] (0,0) -- (4,4) -- (6,4) -- (6,-4) -- (4,-4) -- (0,0);
\draw [help lines] (-6,0) -- (6,0);
\draw [thick][->] (-4,4) -- (-2,2);
\draw [thick] [->](-2,2)--(2,-2);
\draw [thick] (2,-2) -- (4,-4);
\draw [thick][->] (-4,-4) -- (-2,-2);
\draw [thick][->] (-2,-2)--(2,2);
\draw [thick] (2,2) -- (4,4);
\draw[color=white, fill=white] (1.5,1.4) circle [radius=.2];
\draw[color=white, fill=white] (1.5,-1.4) circle [radius=.2];
\draw[color=white, fill=white] (-3.5,2.5) circle [radius=.2];
\draw[color=white, fill=white] (-3.5,-2.5) circle [radius=.2];
\draw[color=white, fill=white] (3,2) circle [radius=.2];
\draw[color=white, fill=white] (3,-2) circle [radius=.2];
\node at (3.5,3) {$\Sigma_{1}$};
\node at (-3.5,3) {$\Sigma_{2}$};
\node at (-3.5,-3) {$\Sigma_{3}$};
\node at (3.5,-3) {$\Sigma_{4}$};
\node [right] at (6,0) {$\Re{z}$};
\draw [fill] (0,0) circle [radius=0.025];
\node [below] at (0,0) {$\xi$};
\node at (1,.4) {$\Omega_{1}$};
\node at (0,1.08) {$\Omega_{2}$};
\node at (-1,.4) {$\Omega_{3}$};
\node at (-1,-.4) {$\Omega_{4}$};
\node at (0,-1.08) {$\Omega_{5}$};
\node at (1,-.4) {$\Omega_{6}$};
\node at (4,1) {$\Wk{2} = \stril[0] {-\dbar R_1 e^{2it \theta}} $ };
\node at (-4,1) {$\Wk{2} = \striu[0]{ -\dbar R_3 e^{-2it \theta}} $ };
\node at (-4,-1) {$\Wk{2} = \stril[0]{ \dbar R_4 e^{2it \theta}} $ };
\node at (4,-1) {$\Wk{2} = \striu[0]{ \dbar R_6 e^{-2it \theta }} $};
\end{tikzpicture}
\caption{The contours $\Sigma_{k}$ and regions $\Omega_k$ $k=1,\dots,6$ defining the $\bar{\partial}$-relationship for the matrix $\mk{2}$. The support of the $\bar{\partial}$-derivatives, $\dbar \mk{2} = \mk{2} \Wk{2}$, is shaded in gray.}\label{fig:M2def}
\end{figure}

Define the contours 
\begin{equation}\label{Sigma_k}
	\Sigma_k = \xi + e^{i(2k-1)\pi/4}\, \R_+ , \quad k = 1,2,3,4,
\end{equation}	
oriented with increasing real part and denote the six open sectors in $\C$ --- separated by $\R$ and the collection of $\Sigma_k$, $k=1,\dots, 4$ --- by $\Omega_k,\, k=1,\dots,6$ starting with the sector $\Omega_1$ between $[\xi, \infty)$ and $\Sigma_1$ and numbered consecutively continuing counterclockwise, see Figure~\ref{fig:M2def}.
Additionally, let 
\begin{equation}\label{distances}
	\mu = \dist(\poles, \R) 
	\qquad 
	\rho = \min \left\{ \min_{\substack{j,k \in \poles \\ j \neq k}} | z_j - z_k |_\infty, \ \mu \right\}
\end{equation}
be the minimal distance from the discrete spectrum to the real axis (positive by assumption) and the lesser of $\mu$ and the minimal $\infty$-norm distance between points of discrete spectra respectively. 
Let $\indicator \in C_0^\infty(\C,[0,1])$ be supported near the discrete spectrum such that 
\begin{equation}\label{chi prop}
	\indicator(z) = \begin{cases}
		1 & \dist(z, \poles \cup \poles^*) < \mu/3 \\
		0 & \dist(z, \poles \cup \poles^*) > 2\mu/3
	\end{cases}
\end{equation}

Standard practice in the analysis of RHPs dictates that we should extend the first and last terms in each factorization in \eqref{V1} to the right and left sides of the contour respectively. We define these extensions of the off-diagonal entries in \eqref{V1} in the following lemma. 

\begin{lemma}\label{lem:extensions}
It is possible to define functions $R_j: \overline{\Omega}_j \to \C$, $j=1,3,4,6$, with boundary values satsifying
\begin{align*}
	R_1(z) &= \begin{dcases}
		r(z) T(z)^{-2} & z \in (\xi, \infty) \\
		\mathrlap {r(\xi) T_0(\xi)^{-2} (z-\xi)^{-2i \kappa(\xi)}(1-\indicator(z)) }
		\hphantom{ 
		\frac{ r(\xi)}{1+ |r(\xi)|^2} T_0(\xi)^{-2} (z-\xi)^{-2i \kappa(\xi)}
		(1-\indicator(z)) }
		& z \in  \Sigma_1
	\end{dcases} \bigskip \\
	R_3(z) &= \begin{dcases}
		\frac{ r(z)^*}{1+ |r(z)|^2} T_+(z)^{2} & z \in (-\infty, \xi) \\
		\mathrlap {\frac{ r(\xi)^*}{1+ |r(\xi)|^2} 
		T_0(\xi)^{2}(z-\xi)^{2i \kappa(\xi)} (1-\indicator(z)) }
		\hphantom{ 
		\frac{ r(\xi)}{1+ |r(\xi)|^2} T_0(\xi)^{-2} (z-\xi)^{-2i \kappa(\xi)}
		(1-\indicator(z)) }
		& z \in  \Sigma_2
	\end{dcases} \bigskip \\
	R_4(z) &= \begin{dcases}
		\frac{ r(z)}{1+ |r(z)|^2} T_-(z)^{-2} & z \in (-\infty, \xi) \\
		\frac{ r(\xi)}{1+ |r(\xi)|^2} T_0(\xi)^{-2} (z-\xi)^{-2i \kappa(\xi)} 
		(1-\indicator(z)) 
		& z \in  \Sigma_3
	\end{dcases} \bigskip \\
	R_6(z) &= \begin{dcases}
		r(z)^* T(z)^{2} & z \in (\xi, \infty) \\
		\mathrlap{ r(\xi)^* T_0(\xi)^{2} (z-\xi)^{2i \kappa(\xi)}(1-\indicator(z))} 
		\hphantom{ 
		\frac{ r(\xi)}{1+ |r(\xi)|^2} T_0(\xi)^{-2} (z-\xi)^{-2i \kappa(\xi)}
		(1-\indicator(z)) }
		 & z \in  \Sigma_4
	\end{dcases} 
\end{align*}
such that for a fixed constant $c_1 = c_1(\psi_0)$, and a fixed cutoff function $\indicator \in C_0^\infty(\C,[0,1])$ satisfying \eqref{chi prop} we have
\begin{equation}\label{dbar bound}
\begin{gathered}
	\left| \dbar R_j(z) \right| \leq 
	c_1 \indicator(z)
	+ c_1 \left| r' \lp \Re z \rp \right| + c_1 |z-\xi|^{-1/2}, \\
	\dbar R_j(z) = 0 \quad \text{ if } \dist(z, \poles \cup \poles^*) \leq \mu/3.
\end{gathered}
\end{equation}
Moreover, if we set $R: \C \to \C$ by $R(z) \big|_{z \in \Omega_j} = R_j(z)$, (with $R_2(z) = R_5(z) = 0$), the extension can be made to preserve the symmetry $R(z^*)^* = R(z)$.
\end{lemma}

\begin{proof}
Using the constant $T_0(\xi)$ defined in Prop.~\ref{prop:T}, define the functions 
\[
\begin{aligned}
f_1(z) &= r(\xi) T^2(z) T_0(\xi)^{-2} (z- \xi)^{-2 i \kappa(\xi) }
&& z \in \overline{\Omega}_1
\\
f_3(z) &= \frac{r(\xi)^*}{1+ |r(\xi)|^2} T(z)^2 T_0(\xi)^{-2} (z- \xi)^{-2 i \kappa(\xi) }
&& z \in \overline{\Omega}_3,
\end{aligned}
\]

Define, for $z \in \overline{\Omega}_j$, $j=1,3$, the extensions
\[
\begin{gathered}
	R_1(z) = \lb f_1(z) + \lp r( \Re z)  - f_1(z) \rp \cos (2\phi) \rb 
	T(z)^{-2}(1- \indicator(z)), 
	\\
	R_3(z) = \lb f_3(z) + 
	\lp \frac{ r( \Re z)^* }{1+ |r(\Re z)|^2} - f_3(z) \rp 
	\cos (2\phi) \rb  T(z)^{2}  (1- \indicator(z)).
\end{gathered}
\]
The extensions $R_4$ and $R_6$ are defined using part b. of Prop.~\ref{prop:T} and choosing $\indicator(z)$ to respect Schwartz symmetry; we define $R_4 = R_3(z^*)^*$ and $R_6(z) = R_1(z^*)^*$ which preserves the Schwartz reflection symmetry.

We give the rest of the details for $R_1$ only. The other cases are easily inferred. 
Clearly, $R_1(z)$ satisfies the boundary conditions of the lemma as $\cos(2\phi)$ vanishes on $\Sigma_1$ and $\indicator(z)$ is zero on the real axis. 
Since $\dbar = (\partial_x + i \partial_y)/2 = e^{i\phi}( \partial_\rho + i \rho^{-1} \partial_\phi)/2$, we have 	
\[
	\begin{multlined}[.9\textwidth]
	\dbar R_1(z) = 
	- 
	\lb f_1(z) + \lp r( \Re z) - f_1(z) \rp \cos 2\phi \rb T(z)^{-2} 
	\dbar \indicator(z) \\
	+ \lb 
	\frac{1}{2} r' \lp \Re z \rp \cos (2\phi) - 
	i e^{i \phi}\, \frac{ r \lp \Re(z) \rp - f_1(z) }{|z-\xi|} \sin(2\phi)
	\rb T(z)^{-2}(1-\indicator(z)) 
	\end{multlined}
\]	
We arrive at \eqref{dbar bound} by observing that $r(z)$ is bounded (since we are assuming there are no imbedded eigenvalues) and as both $1- \indicator(z)$ and $\indicator'(z)$ are supported away from the discrete spectrum, the poles and zeros of $T(z)$ do not affect the bound. This gives the first two terms in the bound. For the last term we 
write
\[
	\left| r(\Re z) - f_1(z) \right|
	\leq \left| r(\Re z) - r(\xi) \right| + \left| r(\xi) - f_1(z) \right|
\]
and use Cauchy-Scwartz to bound each term as follows:
\begin{gather*}
	\left| r(\Re z) - r(\xi) \right| \leq
	\left| \int_{\xi}^{\Re z} r'(s) ds \right|\leq \|r\|_{H^1(\R)} |z-\xi|^{1/2}
	\shortintertext{and}
	\left| r(\xi) - f_1(z) \right| \leq 
	|r(\xi)|(1+ |r(\xi)|^2)   
	\left| T(z,\xi)^2 - T_0(\xi)^2 (z-\xi)^{2i \kappa(\xi)|} \right| 
	\leq C_\xi \| r \|_{H^1(\R)} |z-\xi|^{1/2}.
\end{gather*}
The last estimate uses \eqref{Tbound} and the fact that $T(z,\xi)$ and $(z-\xi)^{i \kappa(\xi)}$ are bounded functions in a neighborhood of $z=\xi$.
The bound \eqref{dbar bound} for $z \in \Omega_1$ follows immediately. 
\end{proof}

We use the extension in Lemma~\ref{lem:extensions} and the factorized jump matrices in \eqref{V1} to define a new unknown function 
\begin{equation}\label{M2}
	\mk{2}(z) = \begin{cases}
		\mk{1}(z) \stril{-R_1(z) e^{2it\theta} } & z \in \Omega_1 \smallskip \\
		\mk{1}(z) \striu{-R_3(z) e^{-2it\theta} } & z \in \Omega_3 \smallskip \\
		\mk{1}(z) \stril{R_4(z) e^{2it\theta} } & z \in \Omega_4 \smallskip \\
		\mk{1}(z) \striu{R_6(z) e^{-2it\theta} } & z \in \Omega_6 \smallskip \\
		\mk{1}(z) & z \in \Omega_2 \cup \Omega_5
	\end{cases}
\end{equation}
Let $\Sk{2} = \bigcup_{j=1}^4 \Sigma_k$. It is an immediate consequence of Lemma~\ref{lem:extensions} and RHP~\ref{rhp:M1} that $\mk{2}$ satisfies the following $\dbar$-Riemann-Hilbert problem.

\begin{rhp}[$\dbar$-]\label{rhp:m2}
Find a function $\mk{2}: \C \backslash (\Sk{2} \cup \poles \cup \poles^*) \to SL_2(\C)$ with the following properties. 
\begin{enumerate}[1.]
\item $\mk{2}$ has continuous first partial derivatives in $\C \backslash (\Sk{2} \cup \poles \cup \poles^*)$. 
\item $\mk{2}(z) = I + \bigo{z^{-1}}$ as $z\to \infty$.
\item For $z \in \Sk{2}$, the boundary values satisfy the jump relation $\mk{2}_+(z) = \mk{2}_-(z) \vk{2}(z)$, where
\begin{equation}\label{V2}
	\begin{gathered}
	\vk{2}(z) = I + (1-\indicator(z)) \delta\vk{2}, \\
	\delta \vk{2}(z) = \begin{cases}
		\tril[0]{ r(\xi) T_0(\xi)^{-2} (z-\xi)^{-2i \kappa(\xi)} e^{2it \theta} } 
		 & z \in \Sigma_1 \smallskip \\
		\triu[0]{ \frac{r(\xi)^* T_0(\xi)^{2}}{1+|r(\xi)|^2}  
		(z-\xi)^{2i \kappa(\xi)} e^{-2it \theta} } 
		& z \in \Sigma_2 \smallskip\\
		\tril[0]{ \frac{r(\xi)T_0^{-2}(\xi)} {1+|r(\xi)|^2} 
		(z-\xi)^{-2i \kappa(\xi)} e^{2it \theta} } 
		& z \in \Sigma_3 \smallskip \\
		\triu[0]{ r(\xi)^* T_0(\xi)^{2}  
		(z-\xi)^{2i  \kappa(\xi)} e^{-2it \theta} } & z \in \Sigma_4 
	\end{cases}
	\end{gathered}
\end{equation}

\item For $z \in \C$ we have
\[
	\dbar \mk{2} (z) = \mk{2}(z) \Wk{2}(z)
\]	
where
\begin{equation}\label{W2}
	\Wk{2}(z) = \begin{cases}
		\tril[0]{-\dbar R_1(z) e^{2it\theta} } & z \in \Omega_1 \smallskip \\
		\triu[0]{-\dbar R_3(z) e^{-2it\theta} } & z \in \Omega_3 \smallskip \\
		\tril[0]{\dbar R_4(z) e^{2it\theta} } & z \in \Omega_4 \smallskip \\
		\triu[0]{\dbar R_6(z) e^{-2it\theta} } & z \in \Omega_6 \smallskip \\
		\qquad \quad \mathbf{0} & \textrm{elsewhere}
	\end{cases}
\end{equation}	

\item $\mk{2}(z)$ has simple poles at each $z_k \in \poles$ and $z_k^* \in \poles^*$ at which 
	\begin{equation}\label{residues2}
		\begin{aligned}
			\res_{z_k} \mk{2} &= 
			\begin{dcases}
				\lim_{z \to z_k} \mk{2}
            		\triu[0]{ c_k^{-1} (1/T)'(z_k)^{-2} e^{-2it \theta} } 
					& k \in \Delta_\xi^{-} \\
				\lim_{z \to z_k} \mk{2}
					\tril[0]{ c_k T(z_k)^{-2} e^{2it \theta} } 
					& k \in \Delta_\xi^{+} 
			\end{dcases} \\
            		\res_{z_k^*} \mk{2} &= 
			\begin{dcases}
        		\lim_{z \to z^*_k} \mk{2} 
                    \tril[0]{ -(c_k^*)^{-1} T'(z_k^*)^{-2} e^{2it \theta} } 
                    & k \in \Delta_\xi^{-}  \\
        		\lim_{z \to z^*_k} \mk{2} 
                    \triu[0]{ -c^*_k T(z_k^*)^{2} e^{-2it \theta} } 
                    & k \in \Delta_\xi^{+} 
			\end{dcases}
		\end{aligned}
	\end{equation}
\end{enumerate}
\end{rhp}

\begin{remark}
In the $\dbar$-RHP for $\mk{2}$ above, it is useful to recall how the extensions $R_j(z)$ are defined in Lemma~\ref{lem:extensions}, particularly the second condition in \eqref{dbar bound}. Though \eqref{W2} may seem to suggest that $\mk{2}$ is non-analytic near the points of discrete spectra, the $\dbar$-derivative vanishes in small neighborhoods of each point of discrete spectra so that $\mk{2}$ is analytic in each neighborhood.
\end{remark}

The $\dbar$-Riemann-Hilbert problem~\ref{rhp:m2} is now ideally conditioned for large $t$ asymptotic analysis. It has jump matrices which approach identity point-wise, all residues corresponding to solitons whose speeds differ from the characteristic defined by $\xi$ are exponentially small, and Lemma~\ref{lem:extensions} controls the $\dbar$ derivatives in a manageable way. 
The final two sections construct the solution $\mk{2}$ as follows, 
\begin{enumerate}[(1)]
	\item The $\dbar$ component of $\dbar$-RHP~\ref{rhp:m2} is ignored, and we prove the existence of a solution of the resulting pure Riemann-Hilbert problem and compute its asymptotic expansion.
	\item Conjugating off the solution of the first step, we arrive at a pure $\dbar$ problem which we show has a solution and bound its size. 
\end{enumerate}

Unwinding the series of transformations that led from the $M$ to $\mk{2}$ we recover the solution RHP~\ref{rhp:M} and then from \eqref{recover} we recover a long time asymptotic expansion of the solution $q(x,t)$ of NLS for our class of initial data.

\section{Removing the Riemann-Hilbert component of the solution}\label{sec:models}

In this section we build a solution $\Mrhp$ to the Riemann-Hilbert problem that results from the $\dbar$-RHP for $\mk{2}$ by dropping the $\dbar$ component. Specifically, 
\begin{equation}\label{M2RHP}
	\text{\mbox{ \parbox{.8\textwidth}{
	Let $\mk{2}_{\textsc{rhp}}$ be the solution of the Riemann-Hilbert problem resulting from 
	setting $\Wk{2} \equiv 0$ in $\dbar$-RHP~\ref{rhp:m2}. 
	}}}
\end{equation}

In this section we will prove that the solution $\Mrhp$ exists and construct its asymptotic expansion for large $t$. Before we embark upon this adventure, we first show that if $\Mrhp$ exists, it reduces $\dbar$-RHP~\ref{rhp:m2} to a pure $\dbar$ problem. 

\begin{prop}\label{prop:rhp to dbar} 
Suppose that $\Mrhp$ is a solution of the Riemann-Hilbert problem described in \eqref{M2RHP}, then the ratio 
\begin{equation}\label{m3}
	\mk{3} (z):= \mk{2}(z) \Mrhp(z)^{-1}
\end{equation}
is a continuously differentiable function satisfying the following $\dbar$-problem. 

\begin{DBAR}\label{rhp:m3 dbar}
Find a function $\mk{3} : \C \to SL_2(\C)$ with the following properties.
\begin{enumerate}[1.]
	\item $\mk{3}$ has continuous first partial derivatives in $\C$.
	\item $\mk{3}(z) =I +\bigo{z^{-1}}$ as $z \to \infty$. 
	\item For $z \in \C$, we have 
	\begin{gather}
		\dbar \mk{3} (z) = 
		\mk{3}(z) \Wk{3} 		
	\end{gather}
	where $\Wk{3} := \Mrhp(z) \Wk{2}(z) \Mrhp(z)^{-1}$ and $\Wk{2}$ is as defined in \eqref{W2}.
\end{enumerate}	
\end{DBAR}

\end{prop}

\begin{proof}
Both $\mk{2}$ and $\Mrhp$ are unimodular and approach identity as $z$ tends to infinity. It follows from \eqref{m3} that $\mk{3}$ inherits these properties as well as continuous differentiability in $\C \backslash \Sk{2}$. Since both $\mk{2}$ and $\Mrhp$ satisfy the same jump relation \eqref{V2}, we have 
\[
\begin{aligned}
	{\mk{3}}_-^{-1} \mk{3}_+ 
	&= {\Mrhp}_-(z)  \mk{2}_-(z)^{-1} \mk{2}_+ (z) {\Mrhp}_+(z)^{-1} \\
	&= {\Mrhp}_-(z) \vk{2}(z) \lp {\Mrhp}_-(z) \vk{2}(z) \rp^{-1} = I,
\end{aligned}
\]
from which it follows that $\mk{3}$ and its first partials extend continuously to $\Sk{2}$. 

Both $\mk{2}$ and $\Mrhp$ are analytic in some deleted neighborhood of each point of discrete spectra $z_k$ and satisfy the residue relation \eqref{residues2}. Let $N_k$ denote the constant (in $z$) nilpotent matrix which appears in the left side of \eqref{residues2}, then we have the Laurent expansions 
\begin{gather}\label{pole removal}
\begin{aligned}
	\mk{2} (z) &= 
	C_0 \lb \frac { N_k }{z-z_k} + I \rb  + \bigo{z-z_k}, \\
	\Mrhp(z)^{-1}(z) &= 
	\lb \frac { -N_k }{z-z_k} + I \rb  \widehat{C}_0 + \bigo{z-z_k}, \\
\end{aligned} 
\intertext{where $C_0$ and $\widehat{C}_0$ are the constant terms in the Laurent expansions of $\mk{2}(z)$ and $\Mrhp(z)^{-1}$ respectively. This implies that} 
	\mk{2} (z) \Mrhp(z)^{-1}(z) = \bigo{1},
\end{gather}
\ie, $\mk{3}$ has only removable singularities at each $z_k$. 
The last property follows immediately from the definition of $\mk{3}$, exploiting the fact that $\Mrhp$ has no $\dbar$ component:
\[
	\dbar \mk{3} (z) = \dbar \mk{2} (z) \Mrhp (z)^{-1} = \mk{2} \Wk{2}(z) \Mrhp (z)^{-1} = \mk{3} \Wk{3}(z).
\] 
\end{proof}

\subsection{Constructing the model problems}
We will construct the solution $\Mrhp$ by seeking a solution of the form
\begin{equation}\label{E impdef}
	\Mrhp(z) = \begin{cases}
		E(z) \mout(z)  & |z - \xi | > \mu/2 \\
		E(z) \mxi (z) & |z - \xi | < \mu/2
	\end{cases}
\end{equation}
where $\mout$ and $\mxi$ are models which we construct below, and the error $E(z)$, the solution of a small norm Riemann-Hilbert problem, we will prove exists and bound it asymptotically. 

\subsubsection{The outer model: an N-soliton potential}
\label{sec:outer model}

The matrix $\Mrhp$ is meromorphic away from the contour $\Sk{2}$ on which its boundary values satisfy the jump relation \eqref{V2}. However, at any distance from the saddle point $z = \xi$, the jump is uniformly near identity. Specifically, let $\U_\xi$ denote the open neighborhood
\begin{equation}
	\U_\xi = \{ z \, : \, |z - \xi | < \mu/2 \},
\end{equation}
on which $\Mrhp$ is pole free. Using the spectral bounds \eqref{distances} and \eqref{V2} 
we have 
\begin{equation}\label{V2 bound outside} 
	\| \vk{2} - I \|_{L^\infty(\Sk{2}) } = 
	\bigo{ \rho^{-2}e^{-\sqrt{2} t |z-\xi|^2} },
\end{equation}
which is exponentially small in $\Sk{2} \backslash \U_\xi$, since $|z - \xi| \geq \mu/2$ outside $\U_\xi$. 
This estimate justifies constructing a model solution outside $\U_\xi$ which ignores the jumps completely. 

\begin{prop}\label{prop:outermodel} 
Let $\mout: \C \to SL_2(\C)$ be a meromorphic function such that
\begin{itemize}
	\item $\mout$ is a meromorphic function from $\C \to SL_2(\C)$
	\item $\mout(z) = I + \bigo{z^{-1}}$ as $z \to \infty$.
	\item $\mout$ has simple poles at each $z_k \in \poles$ and $z_k^* \in \poles^*$ satisfying the residue relations in \eqref{residues2} with $\mout$ replacing $\mk{2}$. 
\end{itemize}

These conditions uniquely determine $\mout$.
Moreover, 
\[
	\lim_{z \to \infty} 2i z \mout_{12} (z;x,t) = \psi(x,t;\sigma_d^{\textrm{out}})
\]
where $\psi(x,t;\sigma_d^{(\textrm{out})})$ is the $N$-soliton solution of \eqref{nls} corresponding to the discrete scattering data 
$\sigma_d^{(\textrm{out})} :=  \{ z_k , \widetilde{c}_k(\xi) \}_{k=1}^N$ 
where
\[
	\widetilde{c}_k(\xi) = c_k \exp \lp \frac{i}{\pi} \int_{-\infty}^\xi \log(1+ | r(s)|^2) \frac{ds}{s- z_k} \rp.
\]
\end{prop}

\begin{proof}
This is a simple consequence of the results in Appendix~\ref{app:solitons}. 
The properties required of $\mout$ are equivalent to RHP~\ref{rhp:solitons2} with 
$\Delta = \Delta_\xi^-$ and $\sigma_d = \sigma_d^{(\textrm{out})}$. The uniqueness of solution and asymptotic behavior are consequences of Prop.~\ref{prop:soliton existence} and \eqref{soliton2 recover}.
\end{proof}

\subsubsection{Local model near the saddle point $z = \xi$}
For $z \in \U_\xi$ the bound \eqref{V2 bound outside} gives a point-wise, but not uniform estimate on the decay of the jump $\vk{2}$ to identity. In order to arrive at a uniformly small jump Riemann-Hilbert problem for the function $E$, implicitly defined by \eqref{E impdef} we introduce a different local model $\mxi$ which exactly matches the jumps of $\Mrhp$ on $\Sk{2} \cap \U_\xi$. In order to motivate the model let $\zeta = \zeta(z)$ denote the rescaled local variable 
\begin{equation}\label{zeta coord}
	\zeta = \zeta(z) = 2 \sqrt{t} (z-\xi)  
	\quad \Rightarrow \quad 
	\zeta^2/2 = 2t (z-\xi)^2
\end{equation}
which maps $\U_\xi$ to an expanding neighborhood of $\zeta = 0$. Additionally, let 
\begin{equation}\label{rchi}
	r_\xi := r(\xi) T_0(\xi)^{-2}  e^{2i ( \kappa(\xi) \log( 2\sqrt{t}) - t \xi^2 )}. 
\end{equation}
Then, since $1 - \indicator(z) \equiv 1$ for $z \in \U_\xi$, the jumps of $\Mrhp$ in $\U_\xi$ can be expressed as 
\begin{equation}
	 \vk{2}(z) \Bigg|_{z \in \U_\xi} = 
	 \begin{cases}
	 	\tril{ r_\xi\, \zeta(z)^{-2i \kappa(\xi)} e^{ i \zeta(z)^2/2} }
		& z \in \Sigma_1 \smallskip \\
		\triu{ \frac{ r_\xi^*}{1+ |r_\xi|^2} 
		\zeta(z)^{2i \kappa(\xi)} e^{-i \zeta(z)^2/2}}
		& z \in \Sigma_2 \smallskip \\
		\tril{ \frac{ r_\xi}{1+ |r_\xi|^2} 
		\zeta(z)^{-2i \kappa(\xi)} e^{i \zeta(z)^2/2}}
		& z \in \Sigma_3 \smallskip \\
		\triu{ r_\xi^* \, \zeta(z)^{2i \kappa(\xi)} e^{-i \zeta(z)^2/2} }
		& z \in \Sigma_4,
	\end{cases}
\end{equation}	
which are \emph{exactly} the jumps of the parabolic cylinder model problem \eqref{a3} described in Appendix~\ref{sec:PC model}. Then using \eqref{a5} we define the local model $\mxi$ in \eqref{E impdef} by 
\begin{equation}\label{mxi def}
	\mxi(z) = \mout(z) \mPC(\zeta(z), r_\xi), \qquad z \in \U_\xi,
\end{equation}
which satisfies the jump $\vk{2}$ of $\Mrhp$ as $\mout$ is an analytic and bounded function in $\U_\xi$.

\subsection{The small norm Riemann-Hilbert problem for $E(z)$}
Using the functions $\mout$ and $\mxi$ defined by Prop~\ref{prop:outermodel} and \eqref{mxi def} respectively, \eqref{E impdef} implicitly defines an unknown $E(z)$ which is analytic in $\C \backslash \Sk{E}$, 
\[
\Sk{E} = \partial \U_\xi \cup (\Sk{2} \backslash \U_\xi),
\]
where we orient $\partial \U_\xi$ clockwise. 
It is straightforward to show that $E(z)$ must satisfy the following Riemann-Hilbert problem.
\begin{rhp}\label{rhp:E}
Find a holomorphic function $E: \C \backslash \Sk{E} \to SL_2(\C)$ with the following properties
\begin{enumerate}[1.]
\item $E(z) = I + \bigo{z^{-1}}$ as $z \to \infty$. 
\item For each $z \in \Sk{E}$ the boundary values $E_\pm(z)$ satisfy
$E_+(z) = E_-(z) \vk{E}(z)$ where
\begin{equation}\label{Ejump}
	\vk{E}(z) = \begin{cases}
		\mout(z)(z) \vk{2}(z) \mout(z)^{-1} & z \in \Sk{2} \backslash \U_\xi \\
		\mout(z) \mPC(\zeta(z),r_\xi) \mout(z)^{-1} & z \in \partial \U_\xi
	\end{cases}
\end{equation}
\end{enumerate}
\end{rhp}

Starting from \eqref{Ejump} and using \eqref{V2 bound outside} for $z \in \C \backslash \U_\xi$ and, using \eqref{zeta coord},\eqref{a6} and the boundedness of $\mout$ for $z \in \U_\xi$, one finds that
\begin{equation}\label{VE bound pts}
	\left| V_E(z) - I \right| = \begin{cases}
		\bigo{ \rho^{-2} e^{-\sqrt{2} t |z - \xi|^2} } 
		& z \in \Sk{E} \backslash \U_\xi \\
		\bigo{ t^{-1/2} } & z \in \partial \U_\xi,
	\end{cases}	
\end{equation}	
and it follows that 
\begin{equation}\label{VE bound}
	\| V_E - I \|_{L^{k,p}(\Sk{E})} = \bigo{ t^{-1/2} } 
	\qquad 
	p \in [1,\infty],\ k\geq 0.
\end{equation}

This uniformly vanishing bound on $V_E - I$ establishes RHP~\ref{rhp:E} as a small-norm Riemann-Hilbert problem, for which there is a well known existence and uniqueness theorem \cite{DZ2,DZ4,Zhou}. In fact, we may write
\begin{equation}\label{E form}
	E(z) = I + \frac{1}{2\pi i} \int_{\Sk{E}} \frac{ (I + \eta(s))(V_E(s) - I)} {s-z} ds
\end{equation}
where $\eta \in L^2(\Sk{E})$ is the unique solution of 
\begin{equation}\label{eta def}
	\lp \one - C_{\vk{E}} \rp \eta = C_{\vk{E}}I.
\end{equation}
Here $C_{\vk{E}}: L^2(\vk{E}) \to L^2(\vk{E})$ is the singular integral operator defined by $C_{\vk{E}} f = C_-( f (V_E-I))$ where $C_-$ is the Cauchy projection operator
\[
	C_- f(z) \lim_{z \to \Sk{E}_-} \frac{1}{2\pi i} \int_{\Sk{E} } f(s) \frac{ ds }{s-z}.
\]	
It's well known that $\| C_- \|_{L^2(\Gamma) \to L^2(\Gamma)}$ is bounded for a very large class of contours $\Gamma$ including the class of finite unions of analytic curves with finite intersection which includes $\Sk{E}$. It then follows from \eqref{VE bound} that 
\begin{equation}\label{CV bound}
	\| C_{\vk{E}} \|_{L^2(\Sk{E}) \to L^2(\Sk{E})} = \bigo{ t^{-1/2} },
\end{equation}
which guarantees the existence of the resolvent operator $(\one - C_{\vk{E}})^{-1}$ and thus of both $\eta$ and $E$. 

The existence of the solution $E(z)$ completes the definition of $\Mrhp(z)$ given by \eqref{E impdef} which in turn solves \eqref{M2RHP} and thus also justifies the transformation \eqref{m3} of Proposition~\ref{prop:rhp to dbar} to an unknown $\mk{3}$ which satisfies the pure $\dbar$-Problem~\ref{rhp:m3 dbar}.  

In order to reconstruct the solution $\psi(x,t)$ of \eqref{nls} we need the large $z$ behavior of the solution of RHP~\ref{rhp:M}. This will include the large $z$ expansion of $E$ which we give here. Geometrically expanding $(s-z)^{-1}$ for $z$ large in \eqref{E form}, which is justified by the finiteness of moments in \eqref{VE bound}, we have
\begin{gather}
	E(z) = I + z^{-1} E_1  + \bigo{z^{-2}} 
	\shortintertext{where} 
	E_1 = -\frac{1}{2\pi i} \int_{\Sk{E}} ( I + \eta(s) )(\vk{E}(s) - I) \, ds.
\end{gather}
Then using \eqref{eta def}-\eqref{CV bound} and the bounds on $V_E- I$ in \eqref{VE bound pts}-\eqref{VE bound} we have 
\[
	E_1 = -\frac{1}{2\pi i} \oint_{\partial \U_\xi} \lp V^E(s) - I \rp ds + \bigo{t^{-1}}.
\]
This last integral, using \eqref{Ejump}, \eqref{a6}, \eqref{beta expand def} and \eqref{zeta coord} can be asymptotically computed by residues yielding (recall that $\partial \U_\xi$ is clockwise oriented) to leading order
\begin{equation}\label{E1} 
	E_1 = \frac{t^{-1/2}}{2} \,  
	\mout(\xi) 
	\offdiag{-i \beta_{12}(r_\xi)}{i \beta_{21}(r_\xi)}
	\mout(\xi)^{-1}  
	+ \bigo{t^{-1}}.
\end{equation}	

\section{Analysis of the remaining \texorpdfstring{$\dbar$}{DBAR}-problem\label{sec:dbar} }

$\dbar$-Problem~\ref{rhp:m3 dbar} is equivalent to the integral equation 
\begin{equation}\label{integralequation}
\mk{3}(z)=I+\frac{1}{\pi}\int_{\mathbb{C}} \frac{\bar{\partial}\mk{3}(s)}{s-z}dA(s)=I+\frac{1}{\pi}\int_{\mathbb{C}} \frac{\mk{3}(s) \Wk{3}}{s-z}dA(s),
\end{equation}
where $s=u+iv$.
 
Equation \eqref{integralequation}, can be written using operator notation as 
\begin{equation}\label{operatorform}
(I-S)[\mk{3}(z)]=I,
\end{equation}
where $S$ is the solid Cauchy operator
\begin{equation}\label{solidop}
S[f]=\frac{1}{\pi}\int_{\mathbb{C}} \frac{f\Wk{3}}{s-z}dA(s).
\end{equation}

The goal at this point is to show that $S$ is small in operator norm so that \eqref{operatorform} may be inverted by Neumann series.

\begin{prop}\label{prop:bound on operator S} 
There exists a constant $C$ such that for all $t>0$, the operator \eqref{solidop} satisfies the inequality
\begin{equation}\label{Sinfnorm}
	\|S\|_{L^{\infty} \to L^{\infty}} \leq Ct^{-1/4}.
\end{equation}
\end{prop}

\begin{proof}
We detail the case for matrix functions having support in the region $\Omega_{1}$, the case for the other regions follows similarly.  Let $A\in L^{\infty}(\Omega_{1})$, then from \eqref{W2} and \eqref{rhp:m3 dbar} it follows that
\begin{align*}
|S[A]|&\leq \iint_{\Omega_{1}} \frac{|A\Mrhp(z) \Wk{2}(z) \Mrhp(z)^{-1}|}{|s-z|}dA(s)\\
&\leq \| A\|_{\infty}\|\Mrhp(z) \|_{\infty}\|\Mrhp(z)^{-1} \|_{\infty} \iint_{\Omega_{1}} \frac{|\bar{\partial}R_{1}e^{2it\theta}|}{|s-z|}dA(s),\\
\end{align*}
where we note that $\Mrhp(z) \Wk{2}(z) \Mrhp(z)^{-1}$ is supported away from the poles $z_{k}$ so that $\|\cdot\|_{\infty}=\|\cdot\|_{L^{\infty}(\text{supp}(R_{1}))}$.

From \eqref{dbar bound} we have the inequality

\begin{equation}\nonumber \label{three integrals}
|S[A]| \leq C\left(I_1+I_2+I_3\right),
\end{equation}
where
\begin{gather*}
		I_1= \iint_{\Omega_{1}} \frac{|\indicator(z)|e^{-4tv(u-\xi)}}{|s-z|}dA(s),
		\qquad
		I_2= \iint_{\Omega_{1}} \frac{|r'(u)|e^{-4tv(u-\xi)}}{|s-z|}dA(s),
		\shortintertext{and}
		I_3= \iint_{\Omega_{1}} \frac{|z-\xi|^{-1/2}e^{-4tv(u-\xi)}}{|s-z|}dA(s).
\end{gather*}
As detailed in \ref{sec:dbar calculations}, we see that there exist constants $c_{1}$, $c_{2}$, and $c_{3}$ such that for all $t>0$ we have the bounds
\begin{equation}\nonumber
	|I_1|\leq \frac{c_{1}}{t^{1/4}} \qquad |I_2|\leq \frac{c_{2}}{t^{1/4}} \qquad |I_3|\leq \frac{c_{3}}{t^{1/4}}.
\end{equation}
and the result is proven.

\end{proof}

For sufficiently large $t$ it is possible to invert the operator \eqref{operatorform} by Neumann series. Furthermore, to detail the long-time asymptotic behavior of $\psi(x,t)$ as mentioned in \eqref{recover}, it is necessary to determine the asymptotic behavior of the coefficient of the $\frac{1}{z}$ term in the Laurent expansion of $\mk{3}$. An integral representation of this term is given by the expansion
\begin{equation}\nonumber
\mk{3}(z)=I+\frac{1}{\pi}\int_{\mathbb{C}} \frac{\mk{3}(s) \Wk{3}}{s-z}dA(s)=I-\frac{1}{\pi}\int_{\mathbb{C}} \left(\frac{\mk{3}(s) \Wk{3}}{z} - \frac{s\mk{3}(s) \Wk{3}}{z(s-z)}\right)dA(s). 
\end{equation}
Therefore we seek the asymptotic behavior of 
\begin{equation}
\mk{3}_{1}=\int_{\mathbb{C}}\mk{3}(s) \Wk{3}dA(s),
\end{equation}
as in the following proposition.

\begin{prop}\label{prop:bound M_1} 
For all $t>0$ there exists a constant $c$ such that 
\begin{equation}
|\mk{3}_{1}|\leq ct^{-3/4}.
\end{equation}
\end{prop}
A proof of Proposition~\ref{prop:bound M_1} is detailed in Appendix~\ref{sec:dbar calculations}.

\section{Long Time Asymptotics for NLS}\label{sec:asymptotics}
From equations \eqref{M1}, \eqref{M2}, \eqref{m3}, and \eqref{E impdef} we see that in the region $\Omega_2$ we have the relationship

\begin{equation*}
M=\mk{3}(z)E(z) \mout(z)T^{\sigma_{3}}.
\end{equation*}
We now take the large $z$ expansions of these matrices to see that
\begin{equation*}
M=\left(I + \frac{\mk{3}_{1}}{z} + \cdots\right)\left(I+ \frac{E_1}{z}+ \cdots\right)\left( I+ \frac{\mout_{1}}{z}+\cdots\right)\left( I+ \frac{T_{1}}{z}+\cdots\right),
\end{equation*}
and consequently the coefficient of the $z^{-1}$ in the Laurent expansion of $M$ is given by
\begin{equation}
M_{1}=\mk{3}_{1}+E_1+\mout_{1}+T_{1}.
\end{equation}

From equation \eqref{recover}, we see that to recover the solution to the NLS equation we seek the off-diagonal entry of $M_{1}$ and therefore find that $T_{1}$ will make no contribution due to its diagonal structure. The asymptotic behavior of $\mk{3}_{1}$ is detailed in Section~\ref{sec:dbar} and does not make the dominant contribution to the asymptotic behavior.
The contribution from $\mout_{1}$ will make a dominant contribution to the asymptotics wherever
$t \to \infty$ along any characteristic $x = x_0 + v t$ inside the truncated cone
\[
	x_1 + v_1 t \leq x \leq x_2+ v_2 t, \qquad t \geq 0
\] as described in Proposition~\ref{prop:soliton separation} of Appendix~\ref{app:solitons}. Away from such a trajectory, the contribution will be exponentially small.
As for the contribution from $E_1$, we use \eqref{E1}, \eqref{rchi}, and the fact that $\det(\mout)=1$ to see that 
\begin{equation}\nonumber
2i(E_1)_{12}=t^{-1/2}\mout(\xi)_{11}^{2}\beta_{12}(r_\xi)+t^{-1/2}\mout(\xi)_{12}^{2}\beta_{21}(r_\xi)+ \bigo{t^{-1}},
\end{equation}
where $\displaystyle \beta_{21}=\frac{\kappa}{\beta_{12}}$ is detailed in \eqref{beta expand def}. Away from the soliton trajectories mentioned above, $\mout(\xi)_{12}$ is exponentially small and $\mout(\xi)_{11}$ is exponentially close to $1$ so that the dominant asymptotic behavior is given by $t^{-1/2}\beta_{12}(r_\xi)$ and the main result follows from \eqref{Tbound} and the identity 
$|\Gamma(i\kappa)|^2=|\Gamma(-i\kappa)|^2=\pi(\kappa \sinh(\pi \kappa))^{-1}$.

\appendix

\section{The parabolic cylinder model problem}\label{sec:PC model}


Let $\Sigma_{PC} = \bigcup_{j=1}^4 \Sigma_j$, where $\Sigma_j$ denotes the complex contour
\begin{equation}\label{a1}
	\Sigma_j  = \left\{ \zeta \in \C \,  | \, \arg \zeta = \frac{2j-1}{4}\pi \right\}, \quad j=1,\dots,4,
\end{equation}	
oriented with increasing real part. Let $\Omega_j$, $j=1,\dots 6$ denote the six maximally connected open sectors in $\C \backslash (\Sigma_{PC} \cup \R)$, where $\Omega_1$ denotes the sector abutting the positive real axis from above, the rest labelled sequentially as one encircles the origin in a counterclockwise fashion. Finally, fix $r \in \C$ and let 
\begin{equation}\label{a2.5}
 	\kappa = \kappa(r) := -\frac{1}{2\pi} \log (1 + |r|^2 ).
\end{equation}	
Then consider the following Riemann-Hilbert problem

\begin{figure}[htb]
	\centering
\begin{tikzpicture}
\draw [help lines] (-2,0) -- (2,0);
\draw [thick][->] (-2,2) -- (-1,1);
\draw [thick] [->](-2,2)--(1,-1);
\draw [thick] (-2,2) -- (2,-2);
\draw [thick][->] (-2,-2) -- (-1,-1);
\draw [thick][->] (-2,-2)--(1,1);
\draw [thick] (-2,-2) -- (2,2);
\node at (2.3,2) {$\Sigma_{1}$};
\node at (-2.3,2) {$\Sigma_{2}$};
\node at (-2.3,-2) {$\Sigma_{3}$};
\node at (2.3,-2) {$\Sigma_{4}$};
\node [right] at (2,0) {$\Re{\zeta}$};
\draw [fill] (0,0) circle [radius=0.025];
\node [below] at (0,-0.1) {$0$};
\node at (1,.4) {$\Omega_{1}$};
\node at (0,1.08) {$\Omega_{2}$};
\node at (-1,.4) {$\Omega_{3}$};
\node at (-1,-.4) {$\Omega_{4}$};
\node at (0,-1.08) {$\Omega_{5}$};
\node at (1,-.4) {$\Omega_{6}$};
\end{tikzpicture}
		\caption{The contours $\Sigma_j$ and sectors $\Omega_j$ in the $\zeta$-plane defining RHP \ref{rhp:a1}. 
		\label{fig:a1} 
		}
\end{figure}
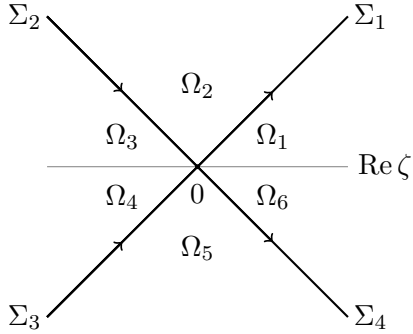

\begin{rhp}[Parabolic Cylinder Model\ ]\label{rhp:a1}
Fix $r \in \C$, find an analytic function $\mk{PC}(\cdot, r): \C \backslash \Sk{PC} \to SL_2(\C)$ such that 
\begin{enumerate}[1.]
\item $\mk{PC}(\zeta,r) = I + \frac{{\mk{PC}}^{(1)}(r) }{\zeta} + \bigo{\zeta^{-2}}$ uniformly as $\zeta \to \infty$.
\item For $\zeta \in \Sk{PC}$,  the continuous boundary values $\mk{PC}_\pm(\zeta,r)$ satisfying the jump relation 
$\mk{PC}_+(\zeta,r) = \mk{PC}_-(\zeta,r) \vk{PC}(\zeta,r)$ where
\begin{equation}\label{a3}
	\vk{PC}(\zeta,r) = \begin{cases}
		\tril{ r \zeta^{-2i\kappa} e^{i \zeta^2/2} } 
		& \arg \zeta = \pi/4 \smallskip \\ 
		\triu{ r^* \zeta^{2i\kappa} e^{-i \zeta^2/2} } 
		& \arg \zeta = -\pi/4 \smallskip \\ 
		\triu{ \frac{r^*}{1+ |r|^2} \zeta^{2i \kappa} e^{-i \zeta^2/2} } 
		& \arg \zeta = 3\pi/4 \smallskip \\ 
		\tril{ \frac{ r}{1+ |r|^2} \zeta^{-2i\kappa} e^{i \zeta^2/2} } 
		& \arg \zeta = -3\pi/4 \\
	\end{cases}
\end{equation}	
\end{enumerate}
\end{rhp}

RHP \ref{rhp:a1} has an explicit solution $\mk{PC}(\zeta,r)$ which is expressed in terms of $D_a(\pm z)$, solutions of the parabolic cylinder equation, $\lp  \pd[2]{}{z} + \lp \frac{1}{2} - \frac{z^2}{2} + a \rp \rp D_a(z) = 0$, as follows:
\begin{gather}\label{a5}
	\mk{PC}(\zeta,r) = \Phi(\zeta,r)  \mathcal{P}(\zeta,r) 
	e^{\frac{i}{4}\zeta^2 \sigma_3} \zeta^{-i \kappa \sigma_3}
	\intertext{where}
	\begin{aligned} \nonumber
	\mathcal{P}(\zeta,r) &= \begin{cases}
		\stril{ -r} & \zeta \in \Omega_1 \medskip \\
		\striu{ \frac{ -r^*}{1+ |r|^2} } & \zeta \in \Omega_3 \medskip \\
		\stril{ \frac{r}{1+ |r|^2} } & \zeta \in \Omega_4 \\
		\striu{r^*} & \zeta \in \Omega_6 \medskip \\
		\quad \ I & \zeta \in \Omega_2 \cup \Omega_5
	\end{cases} 
	\bigskip \\
	\Phi(\zeta,r) &= \begin{cases}
	\begin{pmatrix}
            e^{-\frac{3\pi \kappa}{4}} D_{\scriptscriptstyle i\kappa} \lp e^{\frac{-3i\pi}{4}} \zeta \rp &
            -i {\beta_{12}} e^{\frac{\pi}{4}(\kappa-i) } D_{\scriptscriptstyle -i\kappa-1} \lp
            e^{-\frac{i\pi}{4}} \zeta \rp \\
            i {\beta_{21}} e^{\frac{-3\pi}{4} (\kappa + i) } D_{\scriptscriptstyle i\kappa-1} \lp
            e^{\frac{-3i\pi}{4}} \zeta \rp &
            e^{\frac{\pi\kappa}{4}} D_{\scriptscriptstyle -i\kappa} \lp e^{\frac{-i\pi}{4}} \zeta \rp
	\end{pmatrix} 
	& \zeta \in \C^+ \bigskip \\
	\begin{pmatrix}
            e^{\frac{\pi\kappa}{4}} D_{\scriptscriptstyle i\kappa} \lp e^{\frac{i\pi}{4}} \zeta \rp &
            -i{\beta_{12} } e^{\frac{-3\pi}{4}(\kappa - i) } D_{\scriptscriptstyle -i\kappa-1} \lp
            e^{\frac{3i\pi}{4}} \zeta \rp \\
            i{\beta_{21} } e^{\frac{\pi}{4}(\kappa+i)} 
            D_{\scriptscriptstyle i\kappa-1} \lp e^{\frac{i\pi}{4}} \zeta \rp &
            e^{\frac{-3\pi\kappa}{4}} D_{\scriptscriptstyle -i\kappa} \lp e^{\frac{3i\pi}{4}} \zeta \rp
	\end{pmatrix}
	& \zeta \in \C^-
	\end{cases}  
	\end{aligned} \\
	\intertext{and $\beta_{12}$ and $\beta_{21}$ are the complex constants}
	\label{beta expand def}
	\beta_{12} = \beta_{12}(r) = 
	\frac{ \sqrt{2 \pi} e^{i\pi/4} e^{-\pi \kappa/2} }{ r \Gamma(-i\kappa) }, 
	\qquad 
	\beta_{21} = \beta_{21}(r) = 
	\frac{ -\sqrt{2 \pi} e^{-i\pi/4} e^{-\pi \kappa/2} }{ r^* \Gamma(i\kappa) } 
	= \frac{\kappa}{\beta_{12}}.	
\end{gather}


%

A derivation of this result is given in \cite{DZ3}, a direct verification of the solution in given in the appendix of \cite{JM13}. The essential fact for our needs is the asymptotic behavior of the solution given in the above references, as is easily verified using the well known asymptotic behavior of $D_a(z)$, 
\begin{equation}\label{a6}
	\mk{PC}(\zeta,r) = I + \frac{1}{\zeta} \offdiag{-i\beta_{12}(r)}{i\beta_{21}(r)} + \bigo{\zeta^{-2}}.
\end{equation}

\section{Meromorphic solutions of the NLS Riemann-Hilbert problem \label{app:solitons}}

Here we consider the solutions of the the Riemann-Hilbert problem associated with the NLS equation, RHP~\ref{rhp:M}, for which the reflection coefficient $r(z) \equiv 0$. In this case the unknown function is analytic across the real axis and has isolated poles in the plane, \ie, the solution is meromorphic. The resulting, \emph{reflectionless}, solutions of NLS, $\psi(x,t)$, derived from the solution of the Riemann-Hilbert problem, are (multi-)solitons. Here we give a simple proof of the existence and uniqueness of solutions of this problem and briefly discuss some well known results concerning the asymptotic behavior of these solutions as $t \to \infty.$

Given a finite set of discrete spectra and associated normalization constants, the reflectionless Riemann-Hilbert problem associated to NLS can be stated as follows. 
\begin{rhp}\label{rhp:solitons}
	Given discrete data 
	$\displaystyle{ \data = \left\{ (z_k, c_k) \right\}_{k=1}^N \subset \C^+ \times \C_{*} }$ 
	find a meromorphic function $m : \C \to SL_2(\C)$ with the following properties. 
	\begin{enumerate}[1.]
		\item $m(z)= I + \bigo{z}$  as $z \to \infty$
		\item $m(z)$ is holomorphic in $\C \backslash (\poles \cup \poles^*)$ and has
		simple poles at each point $z_k \in \poles$ and $z_k^* \in \poles^*$ satisfying
		the residue conditions
		\begin{align}\label{b1}
			\res_{z = z_k} m(z) =& \lim_{z \to z_k} m(z) n_k
			\qquad \text{ and } \qquad
			\res_{z = z_k^*} m(z) = \lim_{z \to z_k^*} m(z) \sigma_2 n_k^* \sigma_2 
		\intertext{where $n_k$ is the nilpotent matrix,}
			\nonumber
			n_k &= \tril[0]{ \gamma_k(x,t) } \qquad
			\gamma_k(x,t) := c_k \exp( 2 i ( t z_k^2 + x z_k)).
		\end{align}
	\end{enumerate}
\end{rhp}
It's a direct consequence of the symmetries in RHP~\ref{rhp:solitons} (and more generally in RHP~\ref{rhp:M} ) that any solution of the problem must posses the symmetry $m(z) = \sigma_2 m(z^*)^* \sigma_2$. It follows that any solution of RHP~\ref{rhp:solitons} must admit a partial fraction expansion of the form 
\begin{equation}\label{b2}
	m(z) = I + \sum_{k=1}^N 
	\frac{1}{z-z_k} 
	\begin{pmatrix*}[r]  \alpha_k(x,t) & 0 \\ \beta_k(x,t) & 0  \end{pmatrix*} 
	+ \frac{1}{z-z_k^*}
	\begin{pmatrix*}[r]  0& -\beta_k(x,t)^* \\ 0 & \alpha_k(x,t)^* \end{pmatrix*}
\end{equation}
for coefficients $\alpha_k(x,t), \beta_k(x,t)$ to be determined. 

\begin{prop}\label{prop:soliton existence}
Given data $\data = \{ (z_k, c_k) \}_{k=1}^N \subset \C^+ \times \C_*$ such that $z_j \neq z_k$ for $j \neq k$, RHP~\ref{rhp:solitons} has a unique solution.
\end{prop}

\subsection{Renormalizations of the reflectionless Riemann-Hilbert problem} 
\ \\
The Riemann-Hilbert problem RHP~\ref{rhp:solitons} which encodes the $N$-soliton solutions of \eqref{nls} arises from a particular choice of normalization in the forward scattering step of the IST. Specifically, recalling that $J_1^-(x,t;z)$ and $J_2^+(x,t;z)$ denote the first and second columns respectively of the left and right normalized Jost functions $J^{\pm}(x,t;z)$ of the ZS-AKNS scattering problem, \eqref{laxL}, the matrix $m(z)$ in RHP~\ref{rhp:solitons} is defined for $z \in \C^+$ as
\begin{equation}
	m(z) = m(z;x,t) = \left[ \frac{ J^{-}_1(x,t;z)}{a(z)} \; \Big| \; J^{+}_2(x,t;z) \right] e^{i (t z^2 + x z) \sigma_3},
	\qquad a(z) = \prod_{k=1}^N \lp \frac{ z - z_k}{ z - z_k^*} \rp,
\end{equation}
where $1/a(z)$ is the transmission coefficient of the reflectionless initial data. 
This choice of normalization ensures that for any fixed $t$, $\lim_{x \to + \infty} m(z; x,t) = I$, but is not the only choice available to us. 

Let $\Delta \subseteq \{1, 2, \dots, N\}$ and 
$\nabla = \Delta^c = \{1,\dots,N\} \backslash \Delta$. 
Define 
\begin{equation}\label{a split}
	a_\Delta(z) = \prod_{k \in \Delta} \lp \frac{ z- z_k}{ z -z_k^*} \rp
	\qquad \text{and} \qquad
	a_\nabla(z) = \frac{a(z)}{a_\Delta(z)} 
	= \prod_{k \in \nabla} \lp \frac{ z- z_k}{ z -z_k^*} \rp.
\end{equation}
The renormalization 
\begin{equation}\label{mdelta}
	m^\Delta(z) = m(z) a_\Delta(z)^{\sigma_3} 
	= \left[ \frac{ J^{-}_1(x,t;z)}{a_{\nabla}(z)} \; \Big| \; 
		\frac{J^{+}_2(x,t;z)}{a_\Delta(z)} \right] 
	e^{i (t z^2 + x z) \sigma_3}. 
\end{equation}
then splits the poles between the columns of $m^\Delta(z)$ according to the choice of $\Delta$. 
It's a simple calculation to show that the renormalization $m^\Delta$ satisfies a modified discrete Riemann Hilbert problem.
\begin{rhp}\label{rhp:solitons2}
	Given discrete data 
	$\displaystyle{ \data = \left\{ (z_k, c_k) \right\}_{k=1}^N 
	\subset \C^+ \times \C_{*} }$ 
	and $\Delta  \subseteq \{1,\dots, N\}$ find a meromorphic function 
	$m^\Delta : \C \to SL_2(\C)$ with the following properties. 
	\begin{enumerate}[1.]
		\item $m^\Delta(z)= I + \bigo{z}$  as $z \to \infty$
		\item $m^\Delta(z)$ is holomorphic in $\C \backslash (\poles \cup \poles^*)$ 
		and has simple poles at each point $z_k \in \poles$ and $z_k^* \in \poles^*$
		satisfying the residue conditions
		\begin{gather}\label{b8}
			\res_{z = z_k} m^\Delta(z) = \lim_{z \to z_k} m(z) n_k^{\Delta}
			\qquad \text{ and } \qquad
			\res_{z = z_k^*} m^\Delta(z) = 
			\lim_{z \to z_k^*} m(z) \sigma_2 (n_k^{\Delta})^* \sigma_2 
		\intertext{where $n_k$ is the nilpotent matrix,}
			\nonumber
			n^{\Delta}_k = \begin{cases} 
			\tril[0]{ \gamma_k(x,t) a_\Delta(z_k)^2 } 
			& k \in \nabla \smallskip \\
			\triu[0]{ \gamma_k(x,t)^{-1}  {a_\Delta}'(z_k)^{-2} } 
			& k \in \Delta, 
			\end{cases}
			\quad
			\gamma_k(x,t) := c_k \exp( 2 i ( t z_k^2 + x z_k)),
		\end{gather}
		and $a_\Delta$ is as defined in \eqref{a split}.
	\end{enumerate}
\end{rhp}

As $m^\Delta(z)$ is an explicit transformation of $m(z)$, it follows directly from Prop.~\ref{prop:soliton existence} that RHP~\ref{rhp:solitons2} has a unique solution whenever the poles $z_k \in \poles$ are distinct. 
Moreover, if $\psi_\sol(x,t)=\psi_\sol(x,t; \sigma_d)$ denotes the $N$-soliton solution of \eqref{nls} encoded by RHP~\ref{rhp:solitons}, then using \eqref{recover} and \eqref{mdelta} we have
\begin{equation}\label{soliton2 recover}
	m^\Delta(z) = I + \frac{1}{2i z} 
		\begin{bsmallmatrix} 
		- \int_{x}^\infty |\psi_\sol(s,t)|^2 ds + \sum\limits_{k \in \Delta} 4\Im z_k &
		\psi_\sol(x,t) \\
		\psi_\sol^*(x,t) &
		 \int_{x}^\infty |\psi_\sol(s,t)|^2 ds - \sum\limits_{k \in \Delta} 4\Im z_k 
		\end{bsmallmatrix}
		+ \bigo{z^{-2}}.
\end{equation}	
This shows that each normalization encodes $\psi_\sol$ in the same way. 
The advantage of the nonstandard normalizations is, as we will see below, that by choosing $\Delta$ correctly, other asymptotic limits in which $t \to \infty$ with $-x/2t = \xi$ bounded are under better asymptotic control. 
The new sums appearing on the diagonal entries above, when compared to \eqref{recover}, represent the squared $L^2$ mass of the solitons corresponding to each $z_k$, $k \in \Delta$.

\subsection{Long time behavior of soliton solutions}

If $N=1$, then the scattering data consists of only a single point $\sigma_d = \{ (\xi + i \eta, c_1) \}$. In this case, the algebraic system for $\alpha_1(x,t)$ and $\beta_1(x,t)$ implied by \eqref{b1}-\eqref{b2} is trivial. Using \eqref{recover}, the solution of \eqref{nls}, $\psi(x,t) = -2i \beta_1(x,t)^*$, is given by 
\begin{equation}\label{b5}
\begin{gathered}
	\psi(x,t; \sigma_d) = 2 \eta  \sech \left( 2 \eta (x  + 2 \xi t -x_0) \right)
	e^{-2i (\xi x + (\xi^2 - \eta^2) t)}  e^{-i \phi_0},  \\
	x_0 = \frac{1}{2\eta} \log \left| \frac {c_1}{2 \eta} \right| ,
	\qquad 
	\phi_0 = \frac{\pi}{2} + \arg(c_1),
\end{gathered}
\end{equation}
which is a localized traveling wave of maximum amplitude $2 \Im z_0$ traveling at speed $ - 2 \Re z_0$; the normalization constant $c$  determines the initial location and constant phase shift of the solution. 

For $N>1$ exact formulas for the solution become ungainly, and we will not present them here. However, as is well known, the $N$-soliton solutions undergo elastic collisions and asymptotically separate as $t\to \infty$ into, generically, $N$ single soliton solutions each traveling at speed $-2\Re z_k$, one for each point in the discrete spectra $\{ z_k \}_{k=1}^N$ which define RHP~\ref{rhp:solitons}. 
The exception, of course, is the non-generic case in which two (or more) points of discrete spectra lie on a vertical line $\xi+ i \R$. This can be made precise as follows; for any (possibly degenerate) interval $\Ical = [\xi_1, \xi_2]$  let 
\begin{subequations}\label{soliton parameters}
\begin{equation}
	\poles(\Ical) = \left\{ z_k \in \poles \,:\, \Re z_k \in \Ical \right\}
	\quad \text{and} \quad 
	n(\Ical) = \left| \poles(\Ical) \right| 
\end{equation}
denote the set of point spectra in the vertical strip extending over $\Ical$ and its cardinality respectively; let 
\begin{equation}
	\rho = \rho(\Ical) = \min\limits_{z_k \in \poles \backslash \poles(\Ical) }  
	\Im (z_k) \dist( \Re z_k, \Ical )
\end{equation}	
\end{subequations}

\begin{prop}\label{prop:soliton separation}
Let $\psi(x,t;\data)$ denote the $N$-soliton solution of the NLS equation \eqref{nls} corresponding to discrete scattering data $\data = \{ (z_k , c_k) \}_{k=1}^N  \subset \C^+ \times \C_*$. Fix $x_1, x_2, v_1, v_2 \in \R$ with $x_1 \leq x_2$ and $v_1 \leq v_2$. Let $\Ical = [-v_2/2, -v_1/2]$. Then as $t \to \infty$ along any characteristic $x = x_0 + v t$ inside the truncated cone
\[
	x_1 + v_1 t \leq x \leq x_2+ v_2 t, \qquad t \geq 0
\]
we have
\begin{equation}\label{separation result}
	\left| \psi(x,t; \data)  - \psi (x,t ; \widehat{\sigma}_d) \right| = \bigo{ e^{-4 \rho t} }
\end{equation}
where $\psi(x,t; \widehat{\sigma}_d)$ is the reduced $N(\mathcal{\Ical})$-soliton solution of $NLS$ given by scattering data $\widehat{\sigma}_d = \{ (z_k, \widehat{c}_k ) \, : \, z_k \in \poles (\Ical) \}$ where
\begin{equation}\label{chat}
	\widehat{c}_k = c_k \prod_{ \substack{
		z_j \in \poles \backslash \poles(\Ical) \\
		\Re z_j < -v_2/2 } } 
		\lp \frac{ z_k - z_j}{z_k - z_j^*}  \rp^2
\end{equation}
\end{prop}

\begin{proof}
Let $x = x_0 + v t$ be a characteristic inside the cone and let $\xi = - v/2$ so that $\xi \in \Ical$. Define $\Delta_\xi^\pm$ as in \eqref{index sets} and let
\[ 
a_{\Delta_\xi^-} (z) = \prod_{k \in \Delta_\xi^{-} } \lp \frac{z - z_k}{z - z_k^*} \rp.
\]
Using $a_{\Delta_\xi^-}$ we renormalize the problem as in \eqref{mdelta} by 
defining\footnote{This transformation can be thought of as a reflectionless version of the more general version \eqref{M1} that appears in the analysis of the full problem.} 
$m^{\Delta_\xi^-}(z) = m(z) {a_{\Delta_\xi^-}}(z)^{\sigma_3}$.
The new unknown $m^{\Delta_\xi^-}(z)$ then satisfies RHP~\ref{rhp:solitons2} with 
$\Delta = \Delta_\xi^-$.
The important fact about this choice of normalization is that 
\[
| \gamma_k(x_0 +v t, t) | = |c_k | \exp(-2 x_0 \Im (z_k)) \exp( -4t  \Im(z_k) \Re(z_k - \xi) ) 
\]
which shows that $|\gamma_k|$ grows with $t$  only for those $z_k$ with $k \in \Delta_\xi^-$. The effect of the renormalization $m^{\Delta_\xi^-}$ is to reciprocate these coefficients in the nilpotent matrices defining the residue conditions. Specifically, as $t \to \infty$ along the characteristic,
\[
	\| n^{\Delta_\xi^-}_k \| = 
	\begin{cases}
		\bigo{1} & z_k \in \poles(\Ical) \\
		\bigo{\exp(-4t \rho) } & z_k \in \poles \backslash \poles(\Ical).
	\end{cases}
\] 
This suggest that the poles in $\poles \backslash \poles(\Ical)$ do not meaningfully contribute to the solution. 
Let $\widehat{m}^{\Delta_\xi^-}$ denote the reduced solution of RHP~\ref{rhp:solitons2} with poles only in $\poles(\Ical)$ which results from ignoring the pole conditions at each $z_k \in \poles \backslash \poles(\Ical)$ for $\widehat{m}^{\Delta_\xi^-}$.  
'Un'-renormalizing the solution $\widehat{m}^{\Delta_\xi^-}$ to $\widehat{m}$ (so that all of the poles are in the first column of $\widehat{m}$) one sees that it is defined by the scattering data $\widehat{\sigma}_d = \{ (z_k, \widehat{c}_k ) \, : \, z_k \in \poles (\Ical) \}$  where the $\widehat{c}_k$ are defined by \eqref{chat}.

The residue relations \eqref{b8} satisfied by $m^{\Delta_\xi^-}(z)$ imply that it admits a partial fraction expansion of the form
\begin{equation}\label{pfraction2}
	m^{\Delta_\xi^-}(z) = I 
	+ \sum_{k \in \Delta^+_\xi} 
	\frac{ \begin{psmallmatrix} \alpha_k & 0 \smallskip \\ \beta_k & 0 \end{psmallmatrix} }{z - z_k}
	+ \frac{ \begin{psmallmatrix*}[r] 0 & -\beta_k^* \smallskip \\ 0 & \alpha_k^* \end{psmallmatrix*} }{z - z_k^*}
	+\sum_{k \in \Delta^-_\xi} 
	\frac{ \begin{psmallmatrix} 0 & \beta_k \smallskip \\ 0 & \alpha_k \end{psmallmatrix} }{z - z_k}
	+ \frac{ \begin{psmallmatrix*}[r] \alpha_k^* & 0 \smallskip \\ -\beta_k^* & 0 \end{psmallmatrix*} }{z - z_k^*}
\end{equation}
whose coefficients satisfy the following system of $2N$ equations: 
\ \\
For each $j \in \Delta_\xi^+$:
\begin{subequations}\label{bigsystem}
\begin{gather}
	\begin{aligned}
		\alpha_j 
		+ \gamma_k(x,t) a_{\Delta_\xi^-} (z_j)^2  \lp 
		\sum_{k \in \Delta_\xi^+} \frac{ \beta_k^*} {z_j - z_k^*} 
		- \sum_{k \in \Delta_\xi^-} \frac{ \beta_k} {z_j - z_k}   
		\rp &= 0 \\
		\beta_j^*
		- \frac{\gamma_k(x,t)^*}{ a_{\Delta_\xi^-} (z_j^*)^{2}} \lp
		\sum_{k \in \Delta_\xi^+} \frac{ \alpha_k} {z_j^* - z_k} 
		+ \sum_{k \in \Delta_\xi^-} \frac{ \alpha_k^*} {z_j^* - z_k^*}
		\rp &=  \frac{\gamma_k(x,t)^*}{ a_{\Delta_\xi^-} (z_j^*)^{2}}
	\end{aligned}
	\shortintertext{For each $j \in \Delta_\xi^-$:}	
	\begin{aligned}
		\alpha_j^* 
		+ \frac{{\gamma_j(x,t)^*}^{-1} }{ {a'}_{\Delta_\xi^-}(z_j)^2 } \lp
		- \sum_{k \in \Delta_\xi^+} \frac{\beta_k^*}{z_j^* - z_k^*}
		+ \sum_{k \in \Delta_\xi^-} \frac{\beta_k}{z_j^* - z_k}
		\rp &= 0 \\
		\beta_j 
		- \frac{\gamma_j(x,t)^{-1} }{ {a'}_{\Delta_\xi^-}(z_j)^2 } \lp
		\sum_{k \in \Delta_\xi^+} \frac{ \alpha_k} {z_j - z_k}
		+ \sum_{k \in \Delta_\xi^-} \frac{ \alpha_k^*} {z_j - z_k^*} \rp
		& = \frac{\gamma_j(x,t)^{-1} }{ {a'}_{\Delta_\xi^-}(z_j)^2 }
	\end{aligned}	
\end{gather}	
\end{subequations}

Letting $\eps := \exp(-4 \rho t)$ and rearranging the variables so that the $2N(\Ical)$ equations for the coefficients corresponding to the poles in $\poles(\Ical)$ come first we can write the full system of $2N$ equations in the block matrix form:
\begin{equation}\label{B15}
	\begin{bmatrix}
		\vect{I } + \widehat{\vect{A}} & \vect{A}_{12}   \\   
		\eps \vect{A}_{21}   & 
		\vect{I} +\eps \vect{A}_{22} 
	\end{bmatrix}
	\vect{x} = 
	\begin{bmatrix}
		\vect{b_1} \\ \eps \vect{b_2} 
	\end{bmatrix}
\end{equation}	
where $\vect{x}$ is the vector of $\alpha_k$ and $\beta_k$'s from \eqref{bigsystem}; each of the coefficient blocks $ \widehat{ \vect{A}}$, $\vect{A}_{12}$, $\vect{A}_{21}$, $\vect{A}_{22}$ and the vectors $\vect{b}_1$ and $\vect{b}_2$ are all $\bigo{1}$ and the  upper left $N(\Ical) \!\! \times \! \! N(\Ical)$ block $\widehat {\vect{A}}$ and target vector $\vect{b}_1$ are precisely the data for the linear system corresponding to the the reduced $N(\Ical)$-soliton problem $\widehat{m}^{\Delta_\xi^-}(z)$. The solvability of the soliton problem guaranteed by Prop~\ref{prop:soliton existence} implies that $\widehat{\vect{A}}$ is invertible and thus \eqref{B15} is equivalent to the system
\begin{equation*}
	\lp \vect{I} +  
	\eps \begin{bmatrix}
	(\vect{I} + \vect{\widehat{A}})^{-1} \vect{A}_{12}\vect{A}_{21}  &
	-(\vect{I} +\vect{\widehat{A}})^{-1} \vect{A}_{12}\vect{A}_{22}  \\ 
	\vect{A}_{21}   & \vect{A}_{22}
	\end{bmatrix}
	\rp \vect{x} =  
	\begin{bmatrix}
		 (\vect{I}+ \vect{\widehat{A}})^{-1}( \vect{b_1} - \eps \vect{A}_{12} \vect{b}_2) \\ 
		\eps \vect{b_2} 
	 \end{bmatrix}. 
\end{equation*}	
As all of the coefficients blocks are $\bigo{1}$ the system is near identity and can be expanded asymptotically in $\eps$ which gives 
\[
	\vect{x} = 
	\begin{bmatrix}
		(\vect{I} + \widehat{\vect{A}}\, )^{-1} \vect{b}_1  \\ \vect{0} 
	\end{bmatrix}
	+ \bigo{\eps}.  
\]
Which justifies the claim that the leading order behavior of $m^{\Delta_\xi^-}(z;x,t)$ in the prescribed wedge is given by $\widehat{m}^{\Delta_\xi^-}(z;x,t)$. The result \eqref{separation result} then follows from \eqref{soliton2 recover} and \eqref{pfraction2}.
\end{proof}

%

\begin{proof}[Proof of Proposition~\ref{prop:soliton existence}]
Inserting the partial fraction expansion \eqref{b2} into the residue conditions \eqref{b1} leads to, after some renormalization, the following linear system of equations for $j=1,\dots,N$,
\begin{equation}\label{b3}
	\widehat{\alpha}_j  
	+ \sum_{k=1}^N  \frac{ {\gamma_j}^{1/2}  {\gamma_k^*}^{1/2} }
	{z_j - z_k^*} \widehat{\beta}_k^* 
	= 0,
	\qquad
	\widehat{\beta}_j^*  
	- \sum_{k=1}^N  \frac{ {\gamma_j^*}^{1/2}  \gamma_k^{1/2} }
	{z_j^* - z_k} \widehat{\alpha}_k, 
	= {\gamma_j^*}^{1/2}
\end{equation}
where we've defined the renormalized parameters 
\begin{equation*}
	\widehat{\alpha}_j =  \alpha_j / \gamma_j^{1/2},
	\quad \text{and} \quad
	\widehat{\beta}_j^* = \beta_j^* / {\gamma_j^*}^{1/2},
\end{equation*}
and for brevity we've suppress the $(x,t)$ dependence of $\alpha_j, \beta_j$, and $\gamma_j$.
Letting $\vect{\widehat{\alpha}} = ( \widehat{\alpha}_1, \dots, \widehat{\alpha}_N )^\intercal$,
$\vect{\widehat{\beta}} = ( \widehat{\beta}_1, \dots, \widehat{\beta}_N )^\intercal$,  
$\vect{\gamma}^{1/2} = ( \gamma_1^{1/2} , \dots, \gamma_N^{1/2} )^\intercal$, and $\vect{A}$ be the $N \times N$ matrix with entries 
\[
	\vect{A}_{jk} = \frac{ -i {\gamma_j^*}^{1/2} \gamma_k^{1/2}}{ (z_j^* -z_k) }, \qquad  j,k = 1, \dots,N
\]
the system \eqref{b3} is equivalent to the block matrix equation
\begin{equation}\label{b4}
	\begin{bmatrix*}[r]
	\vect{I}_N & -i\, \vect{A}^* \, \\
	-i \vect{A} & \vect{I}_N\, 
	\end{bmatrix*}
	\begin{bmatrix} \vect{\widehat{\alpha}}  \smallskip \\ \vect{\widehat{\beta}}^* \end{bmatrix} 
	= \begin{bmatrix} 0 \\ {\vect{\gamma}^*}^{1/2} \end{bmatrix}. 
\end{equation}
Note that $\bm{A}^*$ denotes only the complex, not hermitian, conjugate of $\bm{A}$.
Equation \eqref{b4} will have a unique solution if and only if 
\[
	\det  \begin{bmatrix*}[r]
	\vect{I}_N & -i\, \vect{A}^* \, \\
	-i \vect{A} & \vect{I}_N\, 
	\end{bmatrix*} = \det \lp \vect{I}_N + \vect{A} \vect{A}^* \rp \neq 0.
\]

Clearly, $\vect{A}$ is hermitian. Observing also that $\vect{A}$ has the inner product structure
\[
	\vect{A}_{jk} = \int_0^\infty {\gamma_j^*}^{1/2} \gamma_k^{1/2} 
	e^{i (z_k - z_j^*) s} ds 
	= \left\langle {\gamma_j}^{1/2} e^{i z_j s}, \ {\gamma_k}^{1/2} e^{i z_k s} \right\rangle
\]
where the functions $f_j(s) = {\gamma_j}^{1/2} e^{i z_j s}$ are linearly independent in $L^2(\R_+)$ since $z_j \neq z_k$ by assumption. It follows that $\vect{A}$ is positive definite. 
Let $\vect{A}^{1/2}$ denote the unique positive definite square root of $\vect{A}$.
Now the eigenvalues of $\vect{A} \vect{A}^* = \vect{A}^{1/2} \lp \vect{A}^{1/2} \vect{A}^* \rp$  are the same as those of $\vect{A}^{1/2} (\vect{A}^*) \vect{A}^{1/2}$ which is itself positive definite. If we denote these eigenvalues as $\{ \mu_k\}_{k=1}^N \subset \R_+$ then it follows that 
\[
	\det \lp \vect{I}_n + \vect{A} \vect{A}^* \rp = \prod_{k=1}^N (1 + \mu_k) > 0.
\]
This proves the proposition.
\end{proof}

\section{Details of Calculations for the \texorpdfstring{$\dbar$}{DBAR} problem\label{sec:dbar calculations} 
}
\begin{prop}\label{prop:bound on I1} 
There exist constants $c_{1}$, $c_{2}$, and $c_{3}$ such that for all $t>0$ we have the bounds
\begin{equation}\nonumber
	|I_1|\leq \frac{c_{1}}{t^{1/4}} \qquad |I_2|\leq \frac{c_{2}}{t^{1/4}} \qquad |I_3|\leq \frac{c_{3}}{t^{1/4}}.
\end{equation}
\end{prop}

\begin{proof} The calclulations shown here follow those found in \cite{DM}. We will make use of the fact that

\begin{align*}
\left\Vert \frac{1}{|s-z|}\right\Vert_{L^{2}(v+\xi,\infty)}&= \left( \int_{v+\xi}^{\infty} \frac{1}{(u-\alpha)^2 + (v-\beta)^2}du\right)^{1/2}\\
&\leq \left( \int_{\R} \frac{1}{(s)^2 + (v-\beta)^2}ds\right)^{1/2}\\
&= \left(\frac{\pi}{|v-\beta|}\right)^{1/2},
\end{align*}
where we recall that $s=u+iv$ and $z=\alpha + i\beta$. Therefore, we see that
\begin{align*}
|I_1| &\leq \int_{0}^{\infty}\int_{v+\xi}^{\infty}  \frac{|\indicator(z)|}{|s-z|} e^{-8tv(u-\xi)} dudv\\
&\leq \int_{0}^{\infty} e^{-tv^2} \int_{v+\xi}^{\infty}  \frac{|\indicator(z)|}{|s-z|}dudv\\
&\leq \int_{0}^{\infty} e^{-tv^2} \|\indicator(z)\|_{L^{2}(v+\xi,\infty)} \cdot \left\Vert \frac{1}{|s-z|}\right\Vert_{L^{2}(v+\xi,\infty)}dv\\
&\leq c_{1} \int_{0}^{\infty} e^{-tv^2} \left(\frac{\pi}{|v-\beta|}\right)^{1/2}dv\\
& =c_{1}\left(\int_{0}^{\beta} \frac{e^{-tv^2}}{\sqrt{\beta - v}}dv + \int_{\beta}^{\infty} \frac{e^{-tv^2}}{\sqrt{v-\beta}}dv\right).
\end{align*}
For the first integral we make the substitution $v=\beta w$ and remark that since $t>0$, $\beta > 0$, and $w>0$ we have the inequality $\displaystyle \sqrt{\beta} e^{-t\beta^2 w^2}=\frac{(t^{1/4}\beta w)^{1/2}}{t^{1/4}w^{1/2}}e^{-(t^{1/4}\beta w)^{2}} \leq ct^{-1/4}w^{-1/2}$, so that
\begin{equation}\nonumber
\int_{0}^{\beta} \frac{e^{-tv^2}}{\sqrt{\beta - v}}dv = \int_{0}^{1} \sqrt{\beta}\frac{e^{-t\beta^2 w^2}}{\sqrt{1-w}}dw
\leq ct^{-1/4}\int_{0}^{1} \frac{1}{\sqrt{w(1-w)}}dw
\leq Ct^{-1/4}.
\end{equation}
Furthermore, for the second integral we make the substitution $w=v-\beta$ to get
\begin{equation}\nonumber
\int_{\beta}^{\infty} \frac{e^{-tv^2}}{\sqrt{v-\beta}}dv \leq \int_{0}^{\infty} \frac{e^{-tw^2}}{\sqrt{w}}dw \leq t^{-1/4}\int_{0}^{\infty} \frac{e^{-s^2}}{\sqrt{s}}ds \leq Ct^{-1/4}.
\end{equation}
The bound for $I_{2}$ is similar to $I_{1}$, remarking that $r\in H^{1,1}(\mathbb{R})$ and thus, 

\begin{align*}
|I_2| &\leq \int_{0}^{\infty} e^{-tv^2} \int_{v+\xi}^{\infty}  \frac{|r'(u)|}{|s-z|}dudv\\
&\leq \int_{0}^{\infty} e^{-tv^2} \|r'(u)\|_{L^{2}(v+\xi,\infty)} \cdot \left\Vert \frac{1}{|s-z|}\right\Vert_{L^{2}(v+\xi,\infty)}dv\\
& \leq \frac{c_{2}}{t^{1/4}}.
\end{align*}

To arrive at the third bound, we begin with the following estimates for $p>2$ and $\frac{1}{p}+\frac{1}{q}=1$:
\begin{align}\label{Lp bound}
\left\Vert \frac{1}{\sqrt{|s-\xi|}}\right\Vert_{L^{p}(v+\xi,\infty)}&=\left( \int_{v+\xi}^{\infty} \left(\frac{1}{(u-\xi)^2 + v^2}\right)^{p/4}du\right)^{1/p}\\ \nonumber
&=\left( \int_{v}^{\infty} \frac{1}{\left(u^2 + v^2\right)^{p/4}}du\right)^{1/p}\\ \nonumber
&=(v^{1/p-1/2})\left( \int_{1}^{\infty} \frac{1}{\left(1 + w^2\right)^{p/4}}dw\right)^{1/p}\\ \nonumber
&\leq cv^{1/p-1/2}, \nonumber
\end{align}
\begin{align*}
\left\Vert \frac{1}{|s-z|}\right\Vert_{L^{q}(v+\xi,\infty)}&=\left( \int_{v+\xi}^{\infty} \frac{1}{\left((u-\alpha)^2 + (v-\beta)^2\right)^{q/2}}du\right)^{1/q}\\
&\leq \left( \int_{\R} \frac{1}{{\left(s^2 + (v-\beta)^2\right)^{q/2}}}ds\right)^{1/2}\\
&\leq c|v-\beta|^{1/q-1}.
\end{align*}
We now apply the above estimates to see that 
\begin{align*}
|I_3| &\leq C\int_{0}^{\infty} e^{-tv^2}\left\Vert \frac{1}{\sqrt{|s-\xi|}}\right\Vert_{L^{p}(v+\xi,\infty)} \left\Vert \frac{1}{|s-z|}\right\Vert_{L^{q}(v+\xi,\infty)} dv\\
&\leq C\int_{0}^{\infty} e^{-tv^2} v^{1/p -1/2}|v-\beta|^{1/q-1} dv\\
&\leq C\left(\int_{0}^{\beta} e^{-tv^2} v^{1/p -1/2}(\beta-v)^{1/q-1} dv +\int_{\beta}^{\infty} e^{-tv^2} v^{1/p -1/2}(v-\beta)^{1/q-1} dv\right).
\end{align*}

For the first integral we again use the substitution $v=\beta w$ and the bound $\displaystyle \sqrt{\beta} e^{-t\beta^2 w^2} \leq ct^{-1/4}w^{-1/2}$, so that

\begin{align*}
\int_{0}^{\beta} e^{-tv^2} v^{1/p -1/2}(\beta-v)^{1/q-1} dv &=\int_{0}^{1} \sqrt{\beta}e^{-t\beta^2 w^2}w^{1/p -1/2}(1-w)^{1/q-1}dw\\
&\leq ct^{-1/4}\int_{0}^{1} w^{1/p -1}(1-w)^{1/q-1}dw\\
&\leq Ct^{-1/4}.
\end{align*}
For the final integral, we use the substitution $v=w+\beta$ as above so that 
\begin{align*}
\int_{\beta}^{\infty} e^{-tv^2} v^{1/p -1/2}(v-\beta)^{1/q-1} dv&=\int_{0}^{\infty} e^{-t(w+\beta)^2} (w+\beta)^{1/p -1/2}w^{1/q-1}dw\\
& \leq \int_{0}^{\infty} e^{-tw^2} w^{-1/2}dw\\
& \leq Ct^{-1/4},
\end{align*}
and the result is confirmed.
\end{proof}

\begin{prop}
For all $t>0$ there exists a constant $c$ such that 
\begin{equation}
|\mk{3}_{1}|\leq ct^{-3/4}.
\end{equation}
\end{prop}
\begin{proof}
The proof given here follows calculations that can be found in \cite{DM}. Let $A$ be supported in the region $\Omega_{1}$ such that $A\in L^{\infty}(\Omega)$. Then, where  
\begin{align*}
|\mk{3}_{1}|&\leq \frac{1}{\pi} \int \int_{\Omega_{1}} |A\Mrhp(z) \Wk{2}(z) \Mrhp(z)^{-1}|dA\\
&\leq \frac{1}{\pi}\| A\|_{\infty}\|\Mrhp(z) \|_{\infty}\|\Mrhp(z)^{-1} \|_{\infty}\int \int_{\Omega_{1}} |\bar{\partial}R_{1}e^{2it\theta}|dA\\
&\leq C\left(\int \int_{\Omega_{1}} |\indicator|e^{-tuv}dA+\int \int_{\Omega_{1}} |r'|e^{-tuv}dA+\int \int_{\Omega_{1}} |z-\xi|^{-1/2}e^{-tuv}dA\right)\\
&\leq C(I_{4}+I_{5}+I_{6})
\end{align*}
where again we note that $\Mrhp(z) \Wk{2}(z) \Mrhp(z)^{-1}$ is supported away from the poles $z_{k}$ so that $\|\cdot\|_{\infty}=\|\cdot\|_{L^{\infty}(\text{supp}(R_{1}))}$.

To bound $I_{4}$ we use the Cauchy-Schwarz inequality on the inner integral as follows:
\begin{align*}
|I_4| &\leq \int_{0}^{\infty}  \|\indicator(z)\|_{L^{2}(v+\xi,\infty)} \left(\int_{v+\xi}^{\infty} e^{-2tuv}du\right)^{1/2}dv\\
&\leq ct^{-1/2}\int_{0}^{\infty} \frac{e^{-tv^2}}{\sqrt{v}}dv
\leq ct^{-3/4}\int_{0}^{\infty} \frac{e^{-w^2}}{\sqrt{w}}dw
\leq \frac{c}{t^{3/4}}.
\end{align*}
The bound for $I_5$ follows in the same manner as for $I_{4}$. Turning to $I_6$ we once again use H\"{o}lder's inequality for $2<p<4$ and the bound \eqref{Lp bound}. Thus,
\begin{align*}
|I_6|&\leq c\int_{0}^{\infty} v^{1/p-1/2} \left(\int_{v+\xi}^{\infty} e^{-qtuv}du\right)^{1/q}dv\\
&\leq ct^{-1/q} \int_{0}^{\infty}v^{2/p-3/2} e^{-tv^2}dv
\leq ct^{-3/4} \int_{0}^{\infty}w^{2/p-3/2} e^{-w^2}dw 
\leq \frac{c}{t^{3/4}},
\end{align*}
where we have used the substitution $w=t^{1/2}v$ and the fact that $\displaystyle -1<\frac{2}{p} -\frac{3}{2}<-\frac{1}{2}$.
\end{proof}

\printbibliography

\end{document}